%% file: stau-plain.tex
\documentclass[12pt]{article}
\usepackage{epsfig}
\usepackage{graphics}
\usepackage{amsmath}
\usepackage{afterpage}
\usepackage{wasysym}

\input{newcom}

\newlength{\dinwidth}
\newlength{\dinmargin}
\begin{document}
\newcommand {\gapprox}
   {\raisebox{-0.7ex}{$\stackrel {\textstyle>}{\sim}$}}
\newcommand {\lapprox}
   {\raisebox{-0.7ex}{$\stackrel {\textstyle<}{\sim}$}}
\def\gsim{\,\lower.25ex\hbox{$\scriptstyle\sim$}\kern-1.30ex%
\raise 0.55ex\hbox{$\scriptstyle >$}\,}
\def\lsim{\,\lower.25ex\hbox{$\scriptstyle\sim$}\kern-1.30ex%
\raise 0.55ex\hbox{$\scriptstyle <$}\,}

%
%
\begin{titlepage}

\noindent
\begin{flushleft}
{\tt DESY 09-124    } \\
{\tt July 2009}                  \\
\end{flushleft}

\vspace{1.0cm}

\begin{center}
\begin{Large}

{\bf Prospects for the study of the $\tilde{\tau}$-system in  SPS1a' at the ILC}
\vspace{1.5cm}

Philip Bechtle$^{1}$, Mikael Berggren$^{1}$, Jenny List$^{1}$, 
Peter Schade$^{1,2}$ and Olga Stempel$^{2}$

\end{Large}

\vspace{.3cm}
1- Deutsches Elektronen Synchrotron DESY\\
   Notkestr. 85, D-22607 Hamburg, Germany 
\vspace{.1cm}\\
2- Institut f\"ur Experimentalphysik, Universit\"at Hamburg\\
   Luruper Chaussee 149, D-22761 Hamburg, Germany
\end{center}

\vspace{1cm}

\begin{abstract}
The prospects for the analysis of $\tau$ channels at the
SUSY benchmark point SPS1a', 
especially from $e^+e^- \rightarrow \tilde{\tau}^+_1 \tilde{\tau}^-_1$ 
and $e^+e^- \rightarrow \tilde{\tau}^+_2 \tilde{\tau}^-_2$, have been studied in full
simulation of the ILD detector concept foreseen for the International Linear Collider. 
All accessible SUSY channels as well as 
all Standard Model backgrounds were generated at a centre-of-mass energy 
$E_{CMS}$=500~$\GeV$, including the beam energy spectrum and beam backgrounds 
expected for nominal ILC beam parameters. 
With an integrated luminosity of 500 fb$^{-1}$ delivered to the experiment,
the masses of the $\tilde{\tau}_1$ and $\tilde{\tau}_2$ 
can be measured to $107.73^{+0.03}_{-0.05}~\GeVcc \pm 1.1 \cdot \delta 
\MXN{1}$
and $183 ^{+11}_{-5}~\GeVcc \pm 18 \cdot \delta  \MXN{1}$, 
respectively, while the true values in SPS1a' are
107.9~$\GeVcc$ and  194.9~$\GeVcc$, respectively.
This corresponds to $\delta \mstone / \mstone \sim 0.1 ~\%$ and 
$\delta \msttwo / \msttwo \sim 3 ~\%$
with reasonable assumptions on  $\delta \MXN{1}$.
The cross sections for $\tilde{\tau}_1$ and $\tilde{\tau}_2$ pair production could be obtained
with a precision of 3.1~\% and 4.2~\%, respectively.
Combining the mass and cross section measurement in the
$\sttwo$ channel allows to determine the LSP mass with
a relative error of 1.7~\%, assuming a known $\stau$ mixing angle.
In $\stone \rightarrow \tau \tilde{\chi}^0_1$ decays, the $\tau$ polarisation is 
measured to be $91 \pm 9$~\% and $86 \pm 5$~\% in the
$\pi$ and $\rho$ decay channels of the $\tau$, respectively. The true value in
the simulated SPS1a' sample is 89.6~\%. 
\end{abstract}

\vspace{1.0cm}

\begin{center}
Submitted to {\em Phys. Rev.} {\bf D}
\end{center}

\end{titlepage}

%

%
\newpage
\section{Introduction}
\label{sec:1}
The present study of the SUSY benchmark point SPS1a'~\cite{bib:sps1ap} at the 
International Linear Collider (ILC), 
which was undertaken as a part of the preparation of a Letter of Intent for the 
ILD detector~\cite{bib:ILD},
focuses on channels with $\tau$ leptons in the final state.
Contrary to previous fast simulation 
studies, e.g. ~\cite{bib:diagonal}, ~\cite{bib:dm1},
~\cite{bib:gudiuli},
it is entirely based on full detector simulation of
SUSY processes (signal as well as background), 
all Standard Model (SM) backgrounds, and ILC machine background.
The main objective of the study was to asses the capabilities
of the ILD detector in processes particularly sensitive
to beam-beam background and detector hermeticity.
It is also a good probe for the particle identification capabilities
and momentum resolution of the detector.
Furthermore, as several fast simulation studies have been performed in the
past on similar models,
the results presented in this work will give
the opportunity to validate these
results under more realistic conditions, and thus
gain confidence on the validity of such studies in
other channels.
Finally, the rich phenomenology of SPS1a' makes it a good
model to underline the advantages of having an electron-positron 
collider that is tunable both in energy and beam polarisation.

Specifically, the process
$e^+e^- \rightarrow \stone^+ \stone^- \rightarrow \tau^+ \XN{1} \tau^- \XN{1}$ has been
studied with the goal to determine the achievable precision on the $\stone$ mass, the 
$\stone$ pair production cross section as well as the $\tau$ polarisation in the 
$\stone$ decay. For the latter, the decay modes $\tau \rightarrow \pi^{\pm} \nu_{\tau}$ 
and $\tau \rightarrow \rho^{\pm} \nu_{\tau} \rightarrow \pi^{\pm} \pi^0 
\nu_{\tau}$ have been considered. In addition, the expected precision on the 
$\sttwo$ mass and the $\sttwo$ pair production cross section have been 
determined in the process $e^+e^- \rightarrow \sttwo^+ \sttwo^-$ 
$\rightarrow \tau^+ \XN{1} \tau^- \XN{1}$.

The SUSY benchmark point SPS1a' features a quite
light mass spectrum in the slepton sector, and heavy
squarks. Bosinos up to $\XN{3}$ (in  $\eeto \XN{1} \XN{3}$)
would be produced at $E_{CMS}=500~\GeV$.
It is a pure mSUGRA model~\cite{bib:msugra}, hence R-parity 
and CP are conserved. The unification scale parameters are:
$M_{1/2} = 250~\GeVcc$, $M_0 = 70~\GeVcc$, $A_0 = -300~\GeVcc$,
$\tan{\beta} = 10$, and  sign$(\mu) = +1$.
The point is certainly not in contradiction with any 
experimental limits~\cite{bib:pmssm}.
In fact, it is quite close to the most likely point
indicated by present electro-weak precision measurements,
if these are interpreted within a CMSSM framework~\cite{bib:ewfit}. 
In the present study, the phenomenology of SPS1a' was evaluated using
SPheno~\cite{bib:spheno} to run the uni\-fi\-ca\-tion scale
model to the electro-weak scale.

The $\stone$ is the next-to-lightest SUSY particle, the NLSP, 
with $\mstone = 107.9 ~\GeVcc$ and
$\MXN{1} = 97.7 ~\GeVcc$, so $\Delta(M) = 10.2 ~\GeVcc$. 
Due to this rather small mass difference the typical signature of $e^+ e^- \rightarrow \stone^+ \stone^-$ events are two acollinear $\tau$ leptons with a maximal energy of $E_{\tau,max} = 42.5 ~\GeV$ (for E$_{CMS}$ = $500 ~\GeV$, and $M_{\tau}=1.777 ~\GeVcc$), plus a significant amount of missing momentum due to the escaping neutralinos.
As SPS1a' is a point with a sizable co-annihilation contribution
to the dark matter relic density, $\mstone$ is an especially
important quantity to determine. This is usually done by measuring the upper endpoint of the energy spectrum of the $\tau$ leptons from the $\stone$ decay, which is equal to $E_{\tau,max}$. If also the minimal allowed $\tau$ energy $E_{\tau,min}$ can be determined, both $\mstone$ and $\MXN{1}$ can extracted simultaneously - if not, information on $\MXN{1}$ from other SUSY processes is required. In our case, $E_{\tau,min}$ is as low as $2.6 ~\GeV$. At such low $\tau$ energies, the background from $\gamma \gamma \rightarrow \tau \tau$ events is overwhelming and will pose a problem for the study
of the $\stone$.

The mass of the $\sttwo$ is 194.9~$\GeVcc$, so that $E_{\tau,min} = 35.0 ~\GeV$ and $E_{\tau,max} = 152.2 ~\GeV$. Hence, the $\gamma\gamma$ background is less severe, but instead the background from $\eeto WW \rightarrow \ell \nu \ell \nu$ becomes problematic.
Another consequence of the $\stone$ being the NLSP is that $\tau$ leptons are present in a large fraction of the SUSY decays, so that
SUSY itself will be a major background source for $\tau$ channels.

Another important observable for characterising the $\stau$ system is the $\tau$ polarisation. It depends on both the mixing angle $\theta_{\stau}$ of the chiral $\stau$ eigenstates into the mass eigenstates, as well as on the Higgsino and gaugino components of the $\XN{1}$, since the interaction of gauginos and sfermions conserves chirality, while the Yukawa interaction of the Higgsino flips the chirality. The physics of the $\stau$-sector and especially of the resulting $\tau$ polarisation is discussed in detail in~\cite{bib:diagonal}.


In SPS1a', the $\XN{1}$ is expected to have quite a small
Wino component, so the $\XN{1}$ could be parametrised by a single
Bino-Higgsino mixing angle.
To evaluate the the $\stau$ mixing matrix, one needs to measure
both $\stau$ masses and $\theta_{\stau}$. This can be done
by measuring the cross section and the endpoint of the
energy spectrum of the $\tau$ decay products.
The former depends on $\beta^3=(1-4 \mstau^2/s)^{3/2}$
and $\theta_{\stau}$, while the latter depends on
$\mstau$ and $\MXN{1}$. As both $\stau$ sleptons are accessible,
there are four independent measurements possible to
evaluate four parameters.
However, a more sensitive channel to determine $\theta_{\stau}$ is the mixed production, 
$e^+e^- \rightarrow \stone\sttwo$.
To do so, it is of paramount importance to run the accelerator at a centre-of-mass
energy between the thresholds for $\stone\sttwo$  and $\sttwo\sttwo$
production - in SPS1a' between 303 and 390 $\GeV$ - otherwise
the background from  $\sttwo\sttwo$ becomes too severe.
As the present study is performed at  $E_{CMS}$ = $500 ~\GeV$, we have 
therefore not pursued the determination of the mixing angle.

The outline of the paper is as follows:
We start by discussing the detector model used,
the choice of beam polarisation,
and the methods used to generate the event samples,
followed by a breakdown of the different channels.
In the following section, the method to extract the $\stau$
signal is outlined by introducing the most significant differences
between the signal and the various backgrounds.
We continue with a description of the method used to determine
the endpoint of the $\tau$ spectrum and the cross section.
Because of their different signatures, and different main backgrounds,
the analysis for $\stone$ and $\sttwo$ are
separated into individual subsections.
The subsequent section discusses the determination
of the $\tau$ polarisation, based on the $\tau$ decays into
$\pi$ and $\rho$, respectively.
We conclude with a discussion on possible ameliorations
to be implemented in a future study.

\section{Detector and Simulation}
\label{sec:2}
\begin{figure}[hb]
\begin{center}
\resizebox{0.5\textwidth}{!}{%
   \includegraphics{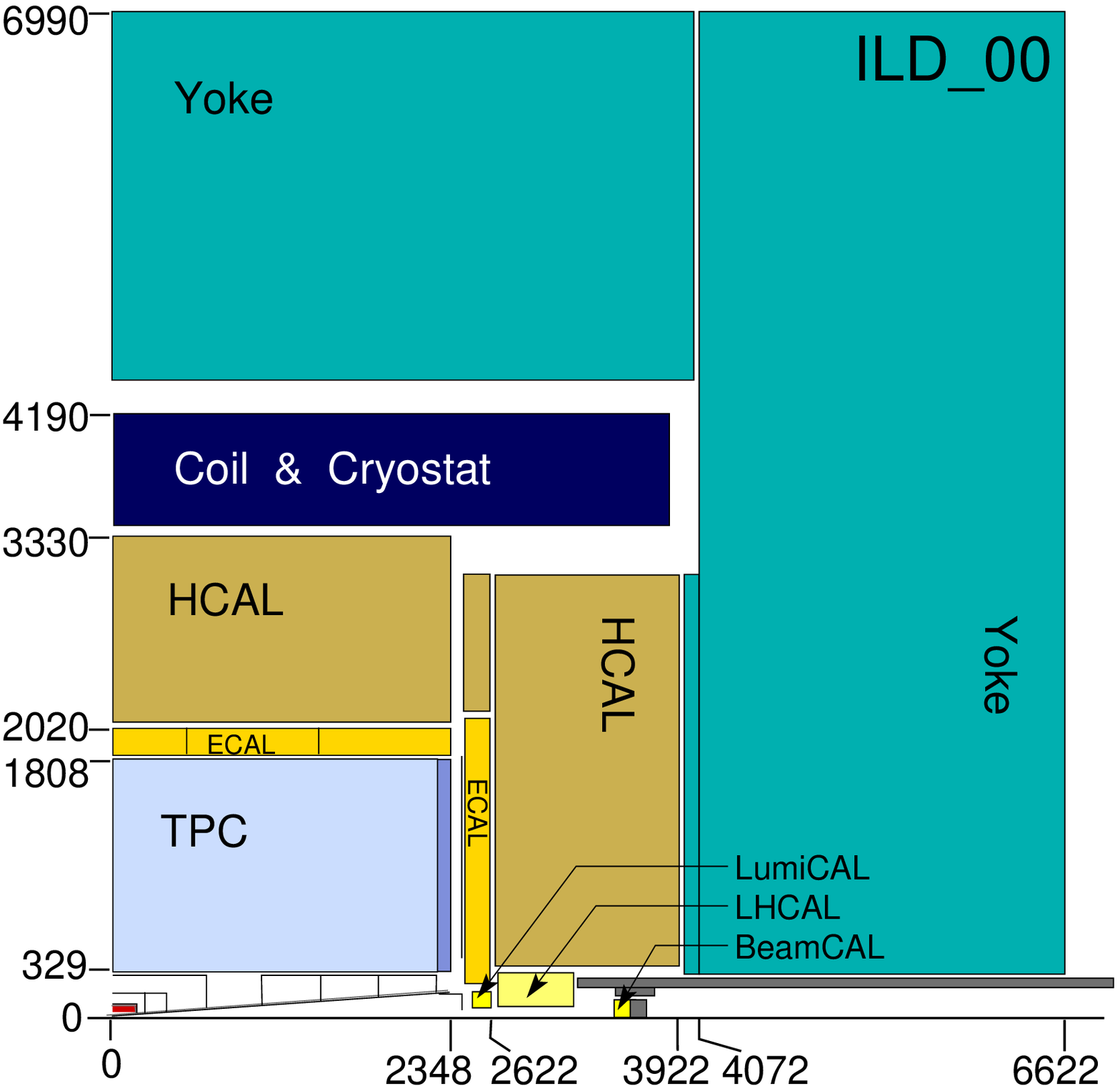}
}
\end{center}
\caption{A quadrant of the ``simulation baseline'' configuration of the
ILD detector. Distances are given in mm. From ~\cite{bib:ILD}.}
\label{fig:ILD}       
\end{figure}

The ILD detector is described in detail in~\cite{bib:ILD}. For the present
study the ``simulation baseline'' detector was used, which is shown
in Fig.~\ref{fig:ILD}.

Of particular importance for the present studies
are the main tracking detector (the TPC), the
main electromagnetic calorimeter (the ECal),
and the low angle calorimeters (the LumiCal,
the LHCal and the BeamCal).


The track finding efficiency, even in high
multiplicity events with overlaid beam background, is
99.5~\% for momenta above 1~\GeVc. 
The transverse momentum resolution
($\Delta(1/P_T)=\Delta(P_T)/P^2_T$) is expected
to be $2.0 \times 10^{-5} ~\GeVc^{-1}$ asymptotically, worsening to 
$9.0 \times 10^{-5} ~\GeVc^{-1}$ at 10 $\GeVc$, and 
to $9.0 \times 10^{-4} ~\GeVc^{-1}$ at 1 $\GeVc$.
In the low angle region, charged tracks will be efficiently
detected down to $\theta=7^\circ$, while the only
region not in the acceptance of the calorimetric system are
the holes in the BeamCal for the beampipes.
Around the outgoing beampipe, the radius of the hole
is 20~mm at $z$=3550~mm, corresponding to 5.6~mrad.
Since the crossing angle of the beams is 14~mrad and the hole for the
incoming beampipe has R=16~mm, 
the lower edge of the acceptance increases to 18.5~mrad at $\phi \approx 180^\circ$.
The ECal is a highly granular SiW sampling calorimeter with a transverse cell size
of 5~mm $\times$ 5~mm and 20 layers. In test-beam measurements with a prototype detector
a resolution of $(16.6\pm 0.1)/\sqrt{E(\mathrm{GeV})}\oplus (1.1 \pm 0.1)\%$ 
has been achieved~\cite{bib:ILD}.
The simulation used here shows a similar resolution.

The \texttt{ILD\_00} detector model was
fully simulated using the 
Geant4-based Mokka~\cite{bib:geant4,bib:mokka}
package. The model not only accounts for the active
elements, but also for support structures, for cables and
cooling systems, and dead regions in the sensitive layers.
In the TPC, the properties of the P5 gas mixture was taken
into account~\cite{bib:tpc}, resulting in a parametrisation of the point 
error depending on both drift distance and local track-pad
angles. Two aspects of the detector, relevant for the
present study, were simulated in less detail: The specific
energy loss in the TPC was estimated by the theoretical Bethe-Bloch
formula, taking into account the actual TPC geometry and read-out granularity, 
to yield an estimate of the separation power between
different particle species~\cite{bib:Mhpc}. This estimate 
was used to simulate
the $dE/dx$-based particle identification
on a track-by-track basis. Furthermore, the response of the BeamCal
to high energy electrons was estimated by tracking the
electron to the BeamCal, and then determining the probability
of detection from a map of the expected energy density
from beamstrahlung pairs, and a parametrisation
of the probability to detect an electron of a given energy
above a given background energy density. Both the map and
the parametrisation were obtained from a separate full simulation
study of the BeamCal alone.

The first means to increase the visibility of the $\stau$ signal above the
background is to determine what beam parameters are the most
favourable.
Because SUSY itself poses a background problem in the $\stau$ analysis,
it is required to run the ILC at the
polarisation that minimises the background.
For 100~\% right  e$^+$ polarisation and 100~\% 
left e$^-$ polarisation ($\mathcal{P}_{beam}(e^+,e^-)$=(+1,$-$1)),
the cross sections for $\XN{2} \XN{2}$ and $\XP{1} \XM{1}$ production are
several 100~fb, and the branching ratios to $\stau$ are above
50~\%. With the opposite polarisation, however, these cross sections
will almost vanish. The SM cross sections are also
reduced for $\mathcal{P}_{beam}$=($-$1,+1),
albeit not so dramatically.
In the case of the $\stone$, an added advantage is
that the production cross section is enhanced
by a factor of 3 for $\mathcal{P}_{beam}$=($-$1,+1) compared to the opposite case.
For the $\sttwo$, the reverse is true, but the gain from the
reduction of the background was found to be the more important feature.
Hence, these channels were studied assuming $\mathcal{P}_{beam}$=($-$0.3,+0.8),
the highest level of polarisation in the advantageous configuration
attainable with the current baseline design of the ILC.

To generate events, the energy spectrum of the ILC beams were simulated first, 
including the effects of both the beamstrahlung and the
energy spread from the main Linac.
With these inputs, SUSY as well as 
Standard Model events were generated by Whizard~\cite{bib:Whizard}.
All SM channels with up to
6 fermions in the final state were simulated. In addition, 
all 8 fermions final states passing the intermediate state 
$t\bar{t}f\bar{f}$ were also generated. 
For channels containing $\tau$ leptons, TAUOLA~\cite{bib:tauola} was used
to generate the $\tau$ decays.
By default, the  helicity of the $\tau$ is only internally generated in 
Whizard,
so an interface between Whizard and TAUOLA was developed in order
to correctly transfer this information between the
two programs. 
In the case of $\gamma\gamma$ events, only 
multi-peripheral diagrams
were included if
the value of $Q^2$ was below 16~$\GeVcc^2$. 
However, other
types of diagrams (VDM, diffraction, etc.) do not produce
events with large missing transverse momentum~\cite{bib:dm1}.
Except for 
the channel $\eeto \gamma\gamma ee \rightarrow \tau\tau e e$,  
a further cut at generator level excluded all $\gamma\gamma$ events
where the
invariant mass of the $f\bar{f}$-pair was below $10 ~\GeVcc$ 
from further treatment.
It should be pointed out that the simulation also includes the
$\gamma$-component of the incoming beams, so that
the $\gamma\gamma$ sample also includes the processes
when one or both of the $\gamma$'s are real.

In addition to these physics channels, the creation of
$e^+ e^-$ pairs due to the beamstrahlung
process were generated, using GuineaPig~\cite{bib:guineapig},
and simulated with Mokka.
Some 125000 such low energetic pairs
are expected to be created in each bunch crossing.
A large fraction of these will leave the detector
through the outgoing beampipe, but nevertheless
the remaining pairs will deposit a large amount
of energy in the BeamCal, and will create a large number
of hits in the inner part of the tracking system.
Approximately one additional charged particle per event will be
detected in the TPC, and a large number of low
energy photons will produce small showers in the
calorimeters, or will convert in the tracking system.
Due to the very large number of particles, this process
cannot be simulated on an event-by-event basis, but
rather a pool of 1000 bunch crossings was 
simulated separately. For each physics event, one such
bunch crossing was selected at random, and overlaid to the 
event at the analysis stage\footnote{This procedure was not
performed for the polarisation measurement.}.
  
The total simulated sample comprised some 13 million
events. The samples generated for the 6 and 8 fermion
channels corresponded to a integrated luminosity of
at least 500 fb$^{-1}$. For the four fermion channels,
the sample sizes for the channels not compatible with $\gamma\gamma$ (i.e.
channels not having an $e^+e^-$-pair in the final state)
corresponded to between 50 and 100 fb$^{-1}$, while the
sample sizes of the channels compatible with  $\gamma\gamma$
corresponded to 0.1 fb$^{-1}$ ($Q^2 < 16~\GeVcc^2$ for both
beam remnants), 1 fb$^{-1}$ ($Q^2 < 16~\GeVcc^2$ for one
beam remnant, $Q^2 > 16~\GeVcc^2$ for the other), or
50~fb$^{-1}$ ($Q^2 > 16~\GeVcc^2$ for both
beam remnants).
Finally, for the two fermion channels, the hadronic and muonic
samples corresponded to 20  fb$^{-1}$, the $\tau^+\tau^-$ sample
to 500 fb$^{-1}$, while the Bhabha channels corresponded to 1~fb$^{-1}$.
(In the Bhabha sample, 
a generator level cut of $\cos{\theta} < 0.96$ and
$\cos{\theta_{acol}} < 0.966$ was imposed).
The total SM background expected at $E_{CMS} = 500~\GeV$ 
for an integrated luminosity of 500 fb$^{-1}$ was
4.92 $\times 10^9$ events, completely dominated by the $\gamma\gamma$ and
Compton scattering processes.
Apart from these two sources, the background was 1.53~$\times 10^8$
events, dominated by 9.94~$\times 10^7$ of $\eeto \ell\ell$
events, mainly Bhabhas.
The SUSY sample was divided into $\stone\stone$, $\sttwo\sttwo$ and
other channels. In all categories, the generated sample corresponded to
at least  500~fb$^{-1}$. The total number of expected events in the
three categories were 7.92 $\times 10^4$, 8.85  $\times 10^3$, and
1.20  $\times 10^4$, respectively.

The simulated events were
reconstructed with MarlinReco~\cite{bib:MarlinReco}.
Tracks in the tracking system were reconstructed using the
Kalman filter method, adopted from DELPHI~\cite{bib:kalman},
and the calorimetric signals were treated using the
particle flow paradigm, implemented with Pandora ~\cite{bib:pandora},
with the PFOid package used for
particle identification.

\section{Mass, cross section and polarisation measurements}
\label{sec:3}

The key characteristics of $\stau$ production and decay, which
single it out from the background, are:
\begin{itemize}
\item only two $\tau$ leptons in the final state
\item large missing energy and momentum
\item high acollinearity, with little correlation to the energy of the
$\tau$ decay products
\item central production
\item no forward-backward asymmetry
\end{itemize}

Different backgrounds dominate for the $\stone$ and the $\sttwo$:
for the $\stone$, the  $\gamma \gamma$ background is important,
while $WW \rightarrow l \nu l \nu$ is less important; the
opposite is true for the $\sttwo$. For the mass measurement,
the SUSY background is not very important, because it is dominated by
$\XPM{1}\XMP{1}$ and $\XN{2}\XN{2}$ production
with cascade decays over $\stau$ sleptons.
In SPS1a', the kinematic limits of
these processes are well below those of both $\stone$ and $\sttwo$ pair production,
so they have little influence on the determination of the
endpoints of the spectra. For the polarisation measurement,
the entire spectrum is needed, and hence the SUSY background becomes
more important. Therefore the selection cuts differ
for the single analyses, but nevertheless two sets of common
initial cuts can be defined, one set to select the signal
topology, and another to reject $\gamma\gamma$ events.

\subsection{Topology selection}
\label{sec:3.1}
The first step in the topology selection was to group
particles into jets.
Jet finding was performed in two ways: for the polarisation analysis,
where no beam induced background was overlaid, the standard 
Durham algorithm was used~\cite{bib:durham}, forced to yield two jets.
In the mass analysis, the overlaid beam induced background
implies that such a method has too low an efficiency due
to extra tracks and clusters from the background. 
In this case,
reconstructed particles to be considered for the jet clustering
were first selected by demanding that all particles should have
an energy exceeding 500 $\MeV$, and that any charged particle
should have at least one hit in the TPC.
To exclude neutrals reconstructed from showers
induced by the beam background, it was demanded that they should 
not be reconstructed from hits in the BeamCal,
nor should the reconstructed starting-point of the
shower be deep in the innermost part of the forward hadronic
calorimeter ($|z| > 3200$~mm  and $ r < 500$~mm).
Then
the algorithm used in DELPHI to find $\tau$ leptons in $\stau$
events was used~\cite{bib:delphimssm}. 
This algorithm - to be applied only after
selecting low charged multiplicity events - goes through all possible
combinations of charged tracks, trying to combine them into
groups with invariant masses of $M_{jet} < 2~\GeVcc$, under the condition that if several
such groupings were possible, the one with the lowest sum of jet-masses should be taken. 
In a second step, neutrals
were added to the charged groups, still respecting the maximal
mass. Any neutrals left over were labelled as belonging
to the ROE group (``Rest Of Event''). 

Events compatible with the $\stau$ topology were then selected 
in both analyses by requiring:

\begin{itemize}
\item exactly two jets
\item less than 10 charged particle candidates
\item vanishing total charge
\item charge of each jet = $\pm 1$
\item invariant jet masses $M_{jet} < 2.5~\GeVcc$ for both jets
\item a total visible energy $E_{vis} < 300~\GeV$
\item a missing mass $M_{miss} > 250~\GeVcc$
\item no particle with momentum above $180 ~\GeVc$
\end{itemize}
Here $M_{miss}$, the invariant mass of the invisible system, is determined
by the difference between the initial $e^+e^-$ system and the
visible system.
After these cuts, 9.23 $\times 10^8$ events remained in the
standard model sample (2.41 $\times 10^5$ non-$\gamma\gamma$).
In the three SUSY subsamples, 6.92 $\times 10^4$,
7.36   $\times 10^3$ and 8.41  $\times 10^3$ events remained
at this stage.

\subsection{ Rejection of $\gamma\gamma$ events, and other SM processes}
\label{sec:3.2}
The characteristics of the 
$\gamma\gamma$ background is the presence of two highly energetic beam remnant 
electrons at low angles, while the rest of the event forms a system of
low energy and mass.
A veto against large energy deposits at small angles
was therefore imposed.
The events that remained did so because the beam remnants escaped
the detector through the incoming or outgoing beampipe.
This limits the missing transverse momentum $P_T$, and implies that the 
visible system consists of two fermions almost
back-to-back in the transverse projection.
In the direction of the incoming beampipe of the other beam,
a much larger deflection of the beam remnant could pass
undetected, so a cut on $P_T$ needs to depend on the azimuthal
angle of the missing momentum.

\begin{figure}[htpb]
\begin{center}
\resizebox{0.31\textwidth}{!}{%
  \includegraphics{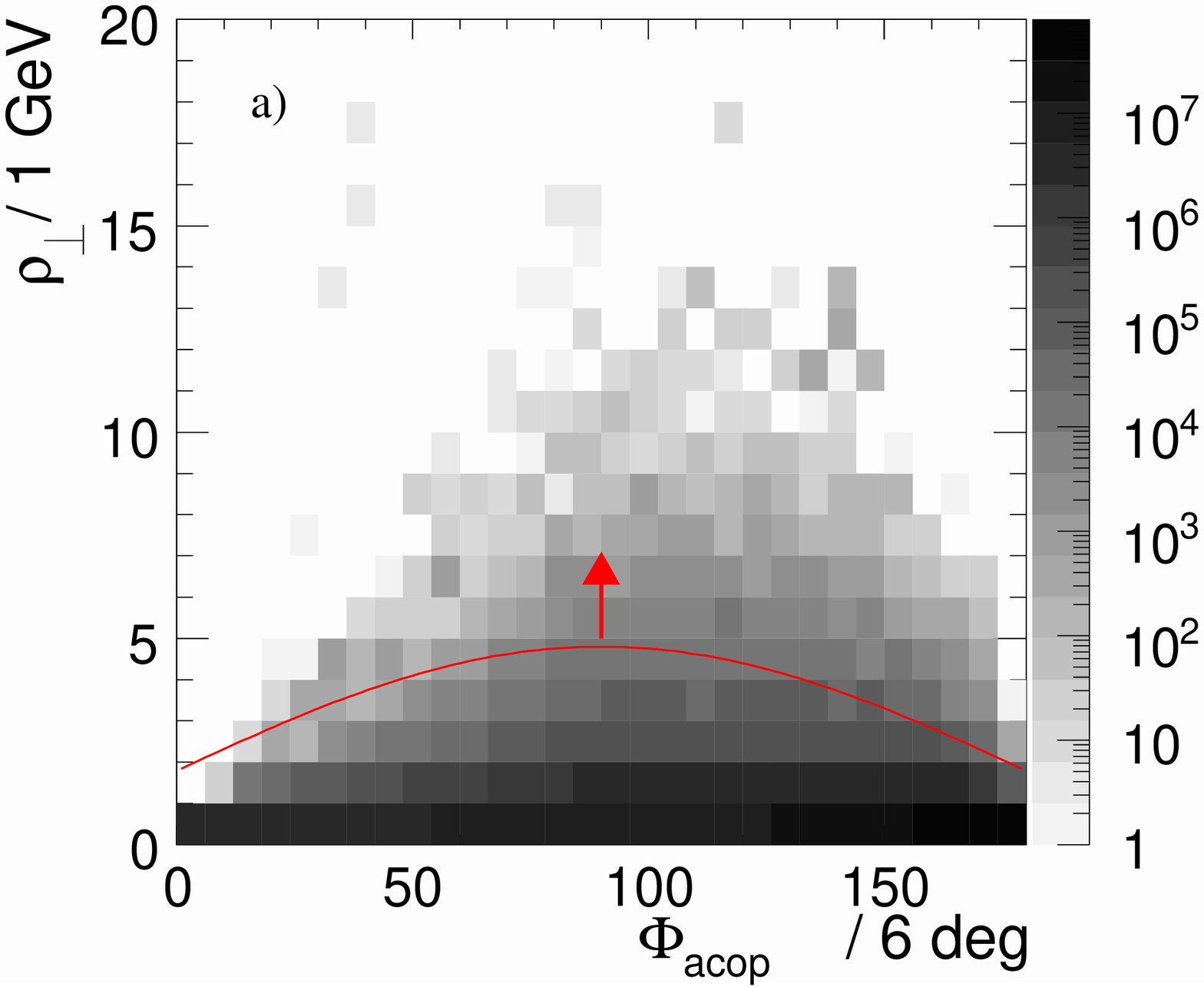}
}
\resizebox{0.31\textwidth}{!}{%
  \includegraphics{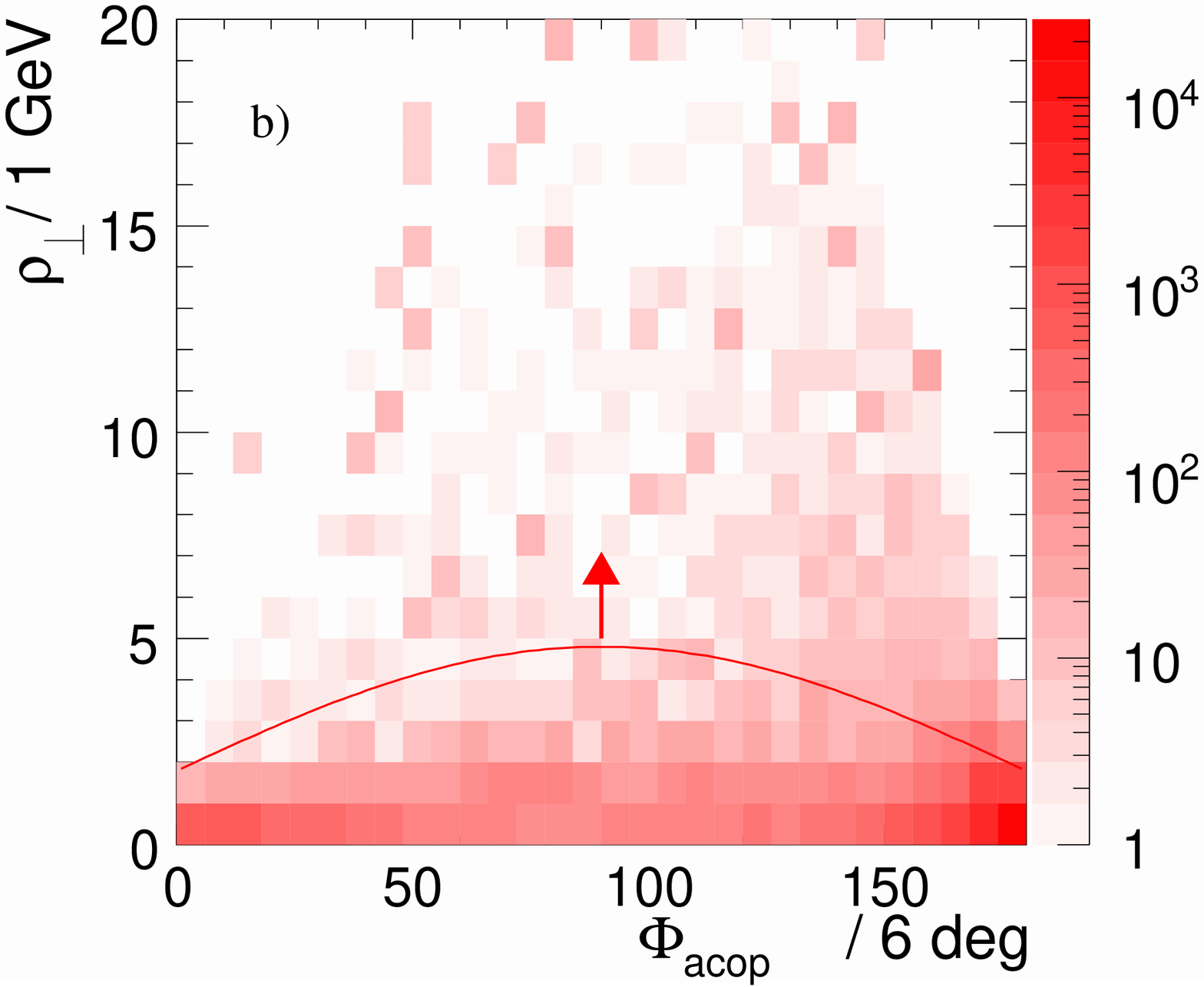}
}
\resizebox{0.31\textwidth}{!}{%
  \includegraphics{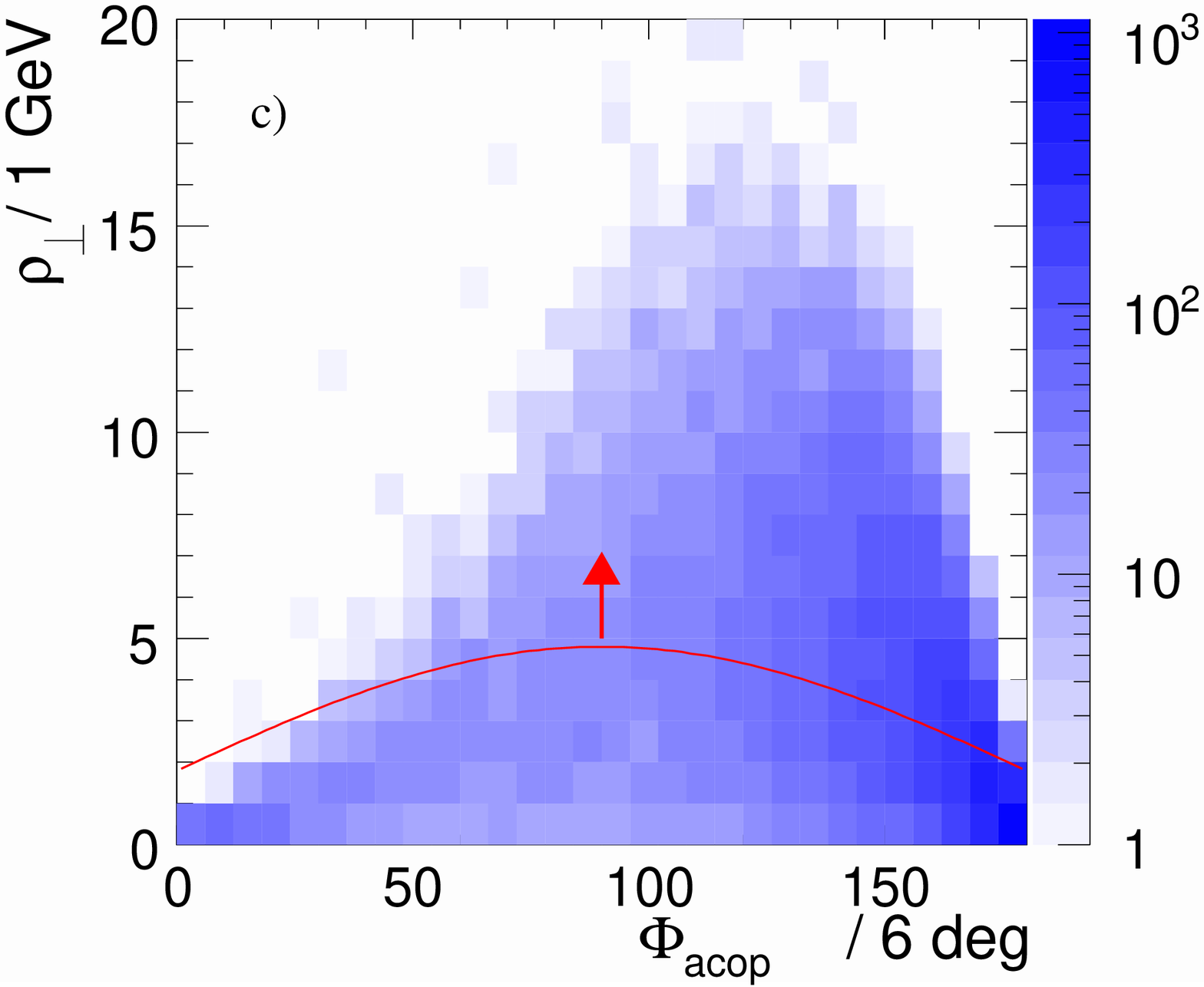}
}
\end{center}
\caption{The distribution of the $\rho_\perp$ variable, defined in the text, versus
the acoplanarity angle. a) $\gamma\gamma$ events, b) Other SM events, c) SUSY signal. 
The selected events are above the solid curve.}
\label{fig:Pt_thrust_rphi}       
\end{figure}

After the low angle veto,  a $\gamma\gamma \rightarrow \tau\tau$
event could only have both high missing $P_T$ and large  $\Phi_{acop}$
(the angle between the two jets
projected to the plane perpendicular to the beam axis)
if the decays of the two $\tau$ leptons were asymmetric:
one of the $\tau$ leptons must have decayed with the visible products
going close to opposite to the $\tau$ direction - giving a
jet at large angle
to the $\tau$, but low energy - while the other must have
done the opposite - giving a jet with small angle to its
parent $\tau$, and large energy.
This property was exploited by the variable $\rho_{\perp}$, ~the scalar
sum of the transverse 
momenta of the jets
w.r.t. the thrust axis, in the projection 
perpendicular to the beam~\cite{bib:alephrho}.
Also $\eeto \tau^+ \tau^-$ events tend to have lower $\rho_\perp$ than the
SUSY signal. Figure ~\ref{fig:Pt_thrust_rphi} shows the
simultaneous distribution of $\rho_\perp$ and $\Phi_{acop}$ for
signal and background events.

\begin{figure}[hb]
\begin{center}
\resizebox{0.5\columnwidth}{!}{%
  \includegraphics{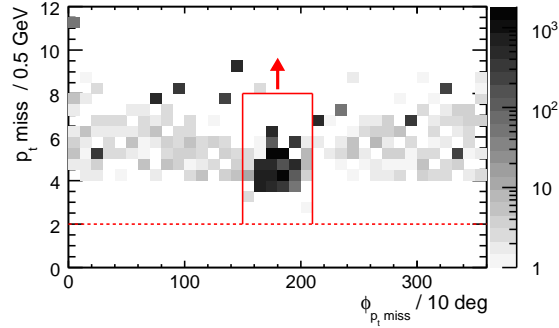}
}
\end{center}
\caption{The distribution of the missing transverse momentum, $p_{_{t}miss}$, as a function
of the azimuthal angle of the missing momentum, $\phi_{p_{t}miss}$. The accumulation of high
$p_{_{t}miss}$ events around $\phi_{p_{t}miss}$= 180$^\circ$ is clearly seen. It corresponds 
to $\gamma\gamma$
events with one beam remnant escaping in the incoming beampipe of the other beam.
Events outside the (red) line were accepted.
}
\label{fig:phi_pt_miss_ver_pt_miss}       
\end{figure}

This lead to the following initial cuts to reduce 
the $\gamma\gamma$ background:
\begin{itemize}
\item $\rho_\perp > (2.7\sin{\Phi_{acop}}+1.8) ~\GeVc$. A
similar 
form of this cut was proposed in~\cite{bib:ulilowdm}.
\item no significant activity in the BeamCal
\item If the azimuthal angle of the
missing momentum, $\phi_{p_{t}miss}$, was within 30$^\circ$ to 
the direction of the
incoming beampipe, the value of the missing transverse momentum
should exceed $8~\GeVc$, see Fig.~\ref{fig:phi_pt_miss_ver_pt_miss}.
\end{itemize}

At this stage, 1.27  $\times 10^6$ events remained, of which
3.82  $\times 10^4$ were  non-$\gamma\gamma$ events.
The reduction of the $\stone\stone$ signal sample, as well
as  of the SUSY background sample, is sizable 
due to these cuts: 2.26  $\times 10^4$ events remained for
$\stone\stone$, while 4.33 $\times 10^3$ remained for the background.
The $\sttwo\sttwo$ sample, on the other hand, was little affected, due
to its higher missing momentum: 6.26 $\times 10^3$  still remained.

In the single analyses, different cuts were used to further
reduce this background to acceptable levels. They were intrinsically
different for the various cases. The polarisation study
relied on particle identification, and this also heavily constrained
the background. In the $\sttwo$ analysis, the signal had high
enough visible energy that the $\gamma\gamma$ background could
easily be reduced.

In the study of the $\stone$ (both mass and polarisation in the $\pi$ channel),
the following cuts were made in addition:
\begin{itemize}
\item
$E_{vis} < 120 ~\GeV$,
\item
$|\cos{\theta_{jet}}| < 0.9$ for both jets,
\item
$\Phi_{acop} > 85^\circ$,
\item
$(E_{jet 1}+E_{jet 2})\sin{\Phi_{acop}}<30~\GeV$.
\end{itemize}

\begin{figure}
\begin{center}
\resizebox{0.31\textwidth}{!}{%
  \includegraphics{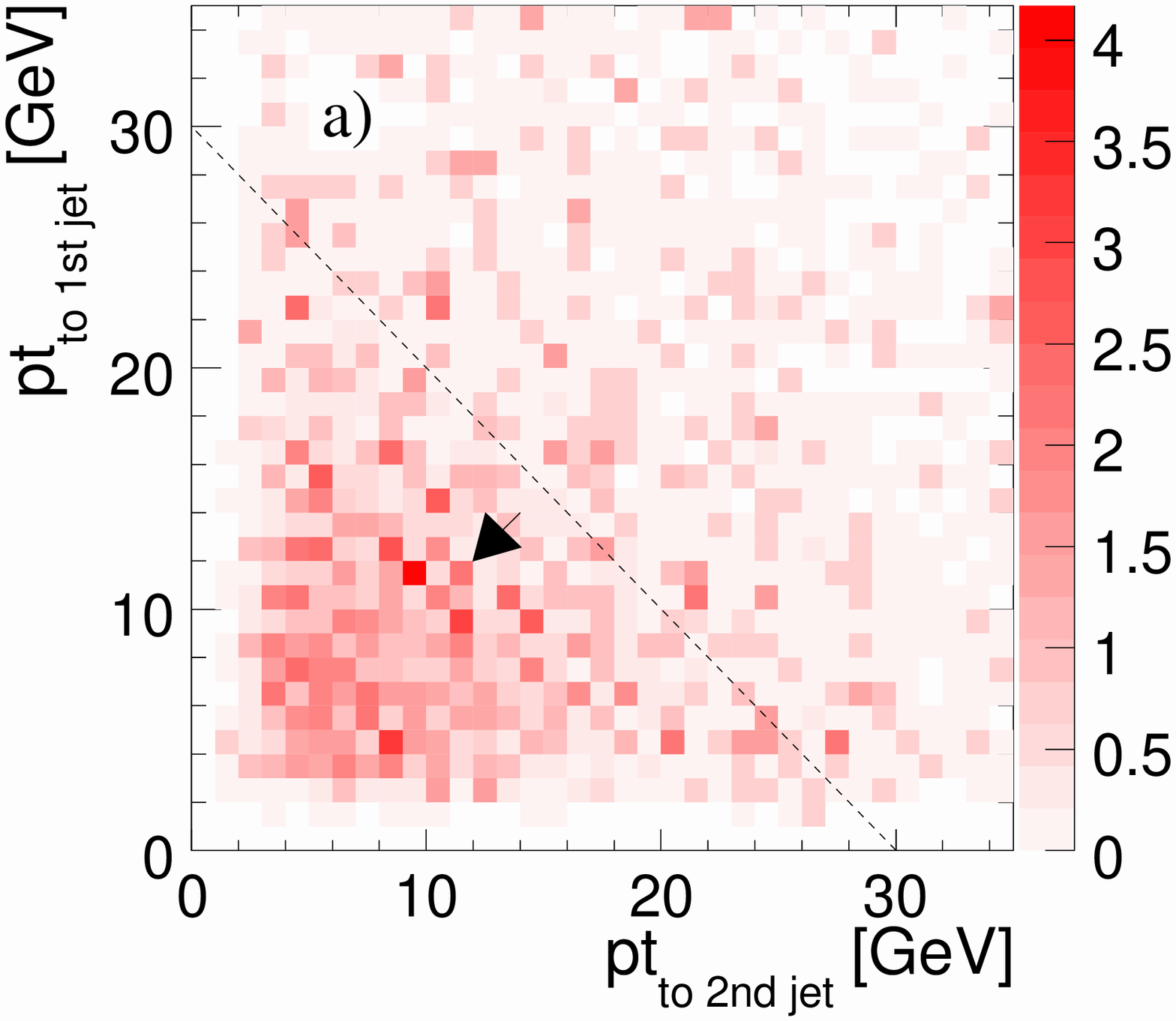}
}
\resizebox{0.275\textwidth}{!}{%
  \includegraphics{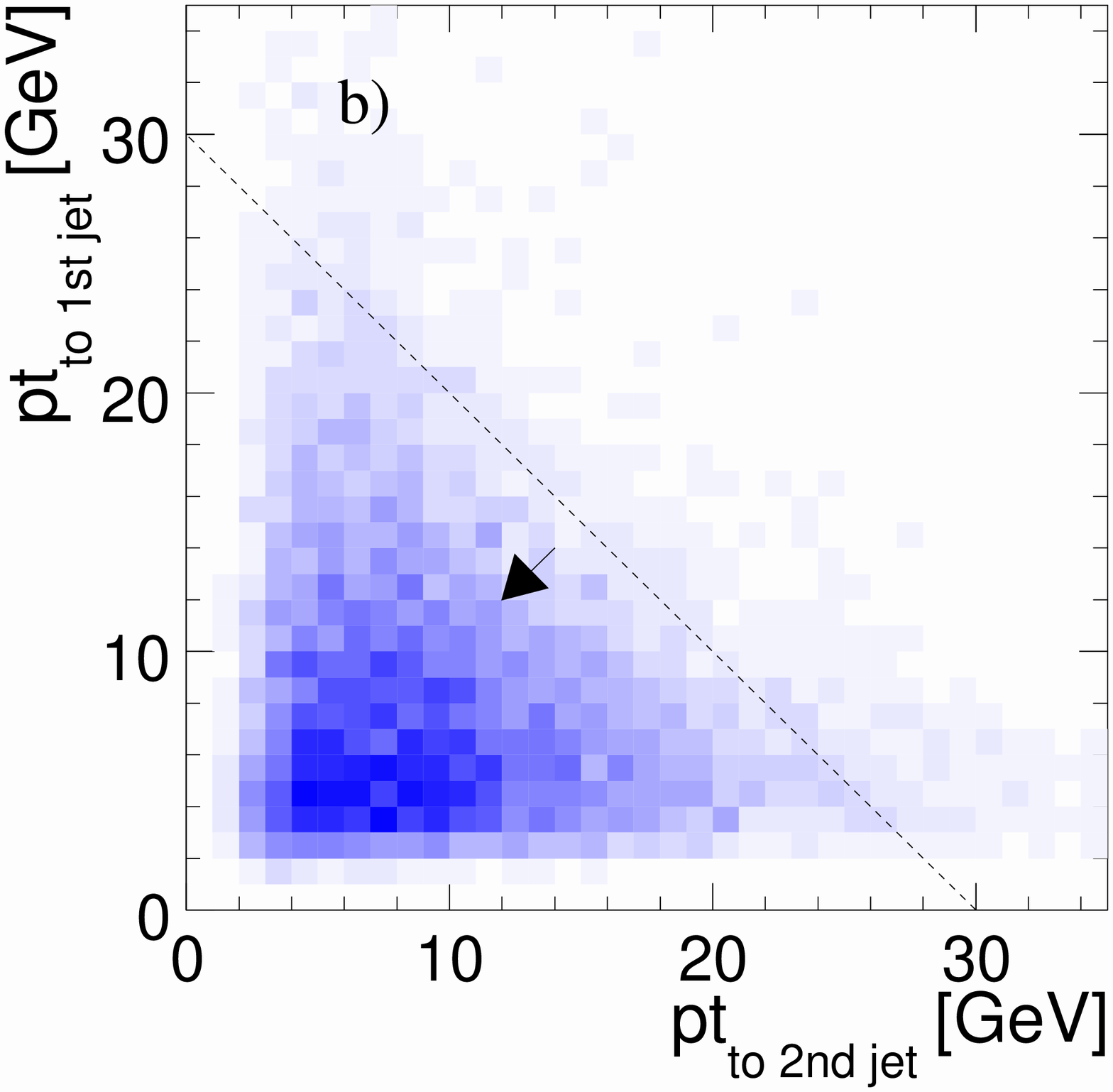}
}
\end{center}
\caption{Transverse momentum of one jet w.r.t. the direction of the other jet. Left: SUSY background.
Right: $\stone$ signal. The cut, given by the dashed line,  corresponds to 
$(E_{jet 1}+E_{jet 2})\sin{\Phi_{acop}}<30$. The selected events are below
the line.}
\label{fig:ptvsother}       
\end{figure}

The cut on $(E_{jet 1}+E_{jet 2})\sin{\Phi_{acop}}$, 
shown in Fig.~\ref{fig:ptvsother}, was effective because
the remaining SUSY background came from decays of heavier sparticles,
and thus tended to be less back-to-back than the signal.
At this stage the SM background was reduced to
9.95  $\times 10^4$ events, of which 1189 were non-$\gamma\gamma$ events.
1.90 $\times 10^4$ $\stone\stone$ events remained, while the SUSY background
was reduced to 2.51 $\times 10^3$ events.  



\subsection{$\stau$ endpoint and cross section}
\label{sec:3.3}
The $\stau$ mass can be extracted from the endpoint of the
$E_{\tau}$ spectrum, which is equal to  $E_{\tau,max}$,
and the $\XN{1}$ mass, known e.g. from a separate analysis
of $\sel$ and $\smu$ decays.
In principle, the maximum of
the spectrum is at $E_{\tau,min}$, so that the $\stau$ channel can
be used to determine $\MXN{1}$ as well, but due to the large
$\gamma \gamma$ background, the turnover point is quite hard
to observe.

In the analysis of both the $\stone$ and $\sttwo$ mass, 
two additional cuts were
applied against the $\gamma\gamma$ background:
The direction of the missing momentum vector was restricted to 
$|\cos{\theta_{miss}}| < 0.8$. In events with more than 
2~$\GeV$ of energy in the ROE group, at least 20~\% of that energy 
should be observed at angles above $30^\circ$ to the beam axis. 
With these cuts, there were 2.22 $\times 10^4$ SM background events left in the
sub-sample selected for the analysis of the $\stone$ properties 
(the `` $\stone$ sample'' in the following). 
At this stage it is a subset of the sample selected for the analysis of the
\sttwo\  properties (the ``\sttwo\ sample''), 
which still contains 1.15 $\times 10^5$ SM events. 
The remaining signal in the two cases was 
1.40 $\times 10^4$ for $\stone\stone$,
and 4242 for $\sttwo\sttwo$.
In addition, the $\stone$ sample contained 214 $\sttwo\sttwo$ events,
and 1559 other SUSY events,
while the  $\sttwo$ sample contained 1.56 $\times 10^4$ $\stone\stone$ events,
and 3033 other SUSY events.

\subsubsection{$\stone$  endpoint and cross section}
\label{sec:3.3.1}
\begin{figure}
\begin{center}
\resizebox{0.55\columnwidth}{!}{%
  \includegraphics{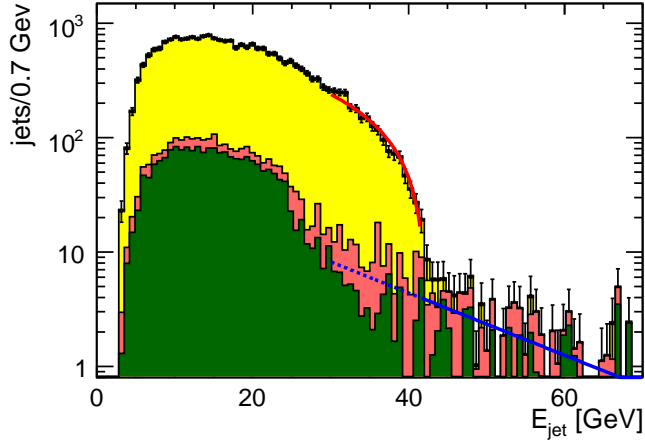}
}
\end{center}
\caption{The jet energy spectrum of events selected in the $\stone$ endpoint analysis, with one entry per jet.
Light grey (yellow) histogram: signal, grey (red) : SM background, 
dark grey (green): SUSY background. The fit to
the background in blue. It is fitted in the signal-free region to the right
(solid portion of the line), and extrapolated into the signal region (dashed).
Fit to total sample: Solid (red) line. The endpoint is at the cross-over of the
two lines.}
\label{fig:st1mass}       
\end{figure}

For the $\stone$ sample, it was finally demanded that the mass of
the visible system, $M_{vis}$,  be above 20 $\GeVcc$ and below 87 $\GeVcc$.
After these cuts, 10244 $\stone$ signal events remained (14.85~\%
efficiency), while the SM background amounts to 323.5 events.
The total SUSY background contained 1029.6 events, including a
contribution of 170.0 $\sttwo$ events. 
Fig.~\ref{fig:st1mass} shows that the endpoint was
almost background free. The turnover point, which is equal to the
minimal $\tau$ energy $E_{\tau,min} = 2.6 ~\GeV$ for most decay
channels, was too distorted by the cuts to be measurable.

The remaining background close to the endpoint - in the range
between 30 and 70 $\GeV$ - was found to be
well described by an exponential, and was fitted in the
signal-free region above 44 $\GeV$. The background fit was
subtracted from the observed spectrum, and the background-subtracted 
spectrum was fitted with a straight line in the range from 30 to 41.5 $\GeV$.
As the dominating background stems from
SM processes, it can be argued that, once ILC data is available to
tune the SM simulation, the background will be known
to a precision much better than what can be determined from the present
simulation.
Therefore, the error on the expectation value of the background 
was assumed to be negligible compared to its Poisson
fluctuations. Hence, the error on the background fit
was not propagated into the statistical errors.
Under these assumptions, the result of the fit 
was $E_{\tau,max} = 42.11^{+0.14}_{-0.12} ~\GeV$.
One notes that this number is not compatible
with the true value ($42.54 ~\GeV$, with mass effects
in the $\tau$ decay included). The difference is 1.1\%, equal to 
3.1 $\sigma$.
It
reflects the fact that a straight line is not quite
adequate to describe the spectrum near the endpoint.
For a final analysis a more sophisticated function would
be needed, either theoretical (including the properties
of the different decay channels and their branching ratios),
or by building MC generated templates.

Fitting $\mstone$ instead of the endpoint position, the result was  
$\mstone = 107.73^{+0.03}_{-0.05}$$~\GeVcc$.
(The true value in SPS1a' is $\mstone = 107.86 ~\GeVcc$.)
However, this result assumes that $\MXN{1}$ is known. At this
model point, the derivative of $\mstone$ w.r.t. $\MXN{1}$ is 1.1,
so the actual error on $\mstone$ is 30 $\MeVcc$ $\oplus$
1.1$\cdot \delta \MXN{1}$. 
In~\cite{bib:ndascthesis}, an analysis of $\smul\smul$ production
using the same fully simulated sample was presented.
The author found  $\delta {\MXN{1}} = 1100~\MeVcc$ from
this channel alone, so using all $\smu$ and $\sel$ channels, one could expect
that $\delta {\MXN{1}} \approx 250~\MeVcc$. This is still a very conservative number, since previous fast simulation
studies of $\smur\smur$ and $\selr\selr$ production in similar, though not
identical scenarios found $\delta \MXN{1} \approx 200~\MeVcc$ and
$\delta {\MXN{1}} \approx 80 ~\MeVcc$, respectively~\cite{bib:ulilowdm}.
In the most optimistic case, the relative error  
$\Delta(\mstone)/\mstone$~is 0.09 \%,
a value still dominated by
$\delta {\MXN{1}}$.

\begin{figure}[hb]
\begin{center}
\resizebox{0.55\columnwidth}{!}{%
  \includegraphics{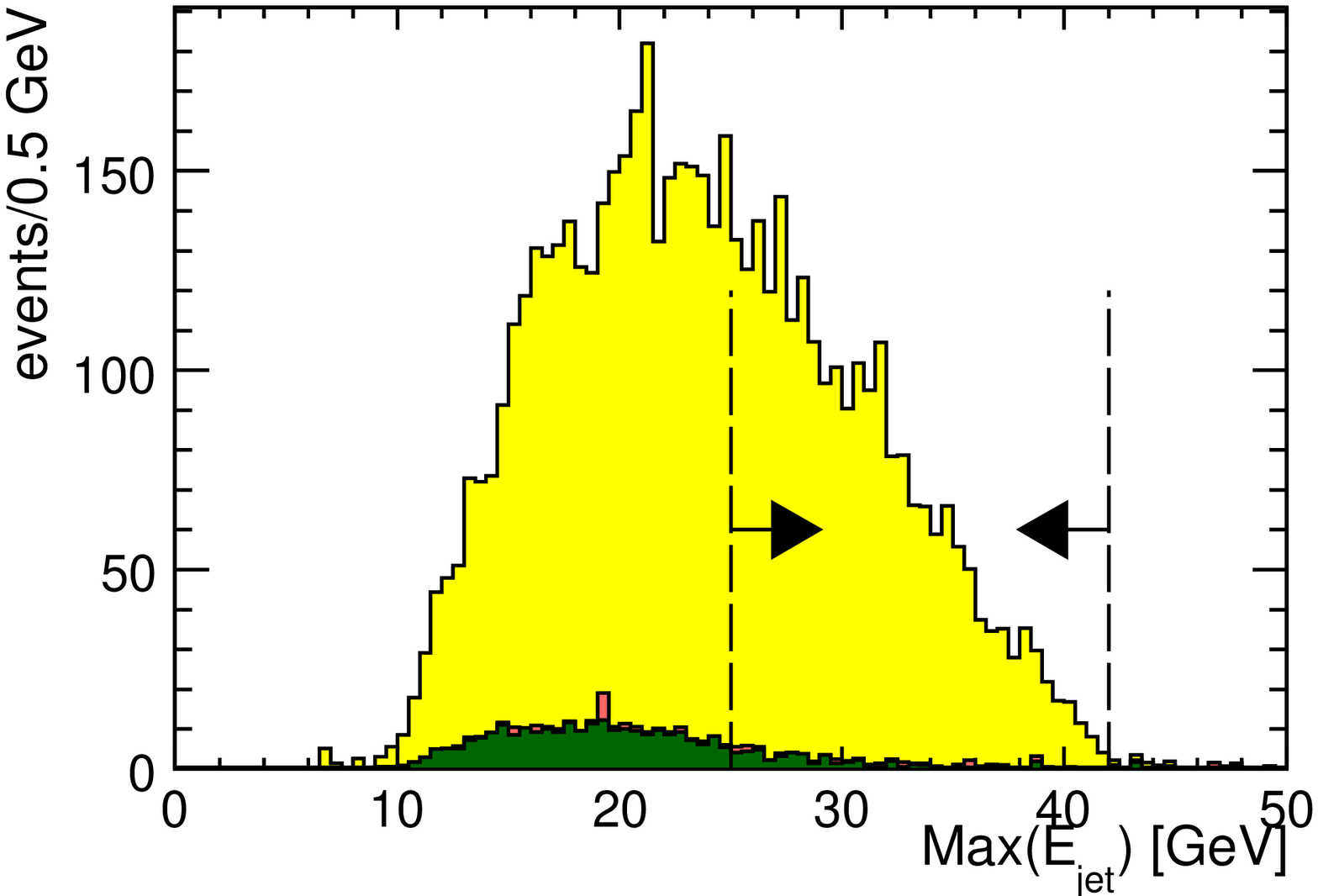}
}
\end{center}
\caption{The spectrum of the highest jet energy of events selected in the $\stone$ cross section analysis, with one entry per event.
Light grey (yellow) histogram: signal, grey (red): SM background, 
dark grey (green): SUSY background.
The cross section was estimated by the number of events having
max($E_{jet}$) between 25 and 42 $\GeV$, as indicated by the vertical lines.}
\label{fig:eps-st1}      
\end{figure}

The ``effective'' cross section, i.e. the cross section for
the beam energy spectrum and the polarisation actually delivered 
by the ILC, was measured as follows: As the main background in this analysis
arose from other SUSY channels, one could not assume that the background
was known with arbitrarily good precision from simulation,
as one can argue for the SM background. It needed either to be estimated
from data, or to be reduced so much that even
a very large uncertainty of its expected level had
little influence on the final number. 
In principle, one could use the fact that the
$\stone$ signal is decreased in  $\mathcal{P}_{beam}=(+0.3,-0.8)$,
while the SUSY background is largely enhanced
to get an equivalent, but almost signal-free, sample.
However, it was quite difficult to reduce the $\stone$ signal
to negligible levels. This would necessitate a thorough
study of the $\XPM{1}$ channel, which was beyond
the scope of the present study.
Instead, another approach was used: by requiring that
at least one jet had an energy above 25 $\GeV$, and none
had an energy above 42 $\GeV$, and that no jet was identified
as a single lepton, the total SUSY background was reduced to
49 events, accompanied by 18 SM events, while 2226 signal events still remained.
The $\max(E_{jet})$ spectrum in this sample is shown in Fig.~\ref{fig:eps-st1}.

Assuming that the uncertainty on the SUSY background is 100~\%,
this yielded a relative  uncertainty on the expected number of
signal events  $\Delta(N_{signal})/N_{signal}=3.1~\%$

This number of events gave an ``effective'' cross section
$\sigma_{eff}$=$N_{signal}/(\mathcal{L} \epsilon_{signal})$=$158.4\pm 4.9$ fb. 
The value is, by construction, equal to the expected one,
and the error only comes from the uncertainty on $N_{signal}$:
The uncertainty on $\mathcal{L}$ is expected to be $<$ 0.1~\% ~\cite{bib:lumi}, i.e.
negligible w.r.t. 3.1~\%. For the present analysis, the error
on $\epsilon_{signal}$ is 2~\%, but this number could be made
significantly smaller with a larger simulated sample.

In the cross section, $\mstone$ only enters via an overall
factor $\beta^3$ : $\sigma_{\stone} = A \times \beta^3/s$, where
$A$ is a function of $\theta_{\stau}$ 
and $\mathcal{P}_{beam}$\footnote{One can note that the maximum cross section 
occurs at $E_{beam}=\sqrt{5/2} \mstone$ i.e. at 170.6~$\GeV$.
Hence, $E_{beam}$= 250~$\GeV$ is not optimal for extracting the
mass from the cross section: the ($\mstone$ independent) $1/s$ decrease 
of the cross section is more important than the $\beta^3$ rise.}.
This gives $\mstone = E_{beam} \sqrt{1-(\sigma s/A)^{2/3}}$, and the  error
on the mass is 
\begin{align*}
\Delta^2(\mstone)= &
(\frac {d\mstone}{d\sigma})^2\Delta(\sigma)^2+
(\frac {d\mstone}{dE})^2 \Delta(E)^2+\\
&(\frac {d\mstone}{dA})^2 \Delta(A)^2
\end{align*}
Considering this at fixed $A$, 
$$
\frac{\Delta(\mstone)}{\mstone} = 
\frac{\sqrt{(\frac{\Delta(\sigma)}{\sigma}\beta^2)^2+
(\frac{\Delta(E)}{E}(3-5\beta^2))^2}}{3(1-\beta^2)}
$$
The design goal for the ILC is $\frac{\Delta(E)}{E} \leq 0.1$\%~\cite{bib:RDR},
so for SPS1a', and the given $E_{beam}$, the second term is negligible,
and $\Delta(\mstone)/\mstone = (\Delta(\sigma)/\sigma)(\beta^2)/3(1-\beta^2)$=
2.1~\%.

Finally, it can be noted that, assuming the background and 
signal selection efficiencies remain the same,
the optimal centre-of-mass energy for determining $\mstone$ from the
cross-section is
250~$\GeV$. With these assumptions, the error on the mass is reduced to
a quarter, i.e. $\Delta(\mstone)/\mstone$ =  0.5~\%.
It is most probable that the background levels would be
substantially lower, since 
there is no SUSY background at $E_{CMS}$ = 250~$\GeV$.
It is possible that the SM background would increase,
due to the change of cuts needed to accommodate the change in
signal characteristics: $E_{\tau,min}$ increases 
to 5.7 $\GeV$,
while $E_{\tau,max}$ decreases to 16.9 $\GeV$.
However, this would only mean trading a poorly known SUSY background for
a well known SM one. 
\subsubsection{$\sttwo$ endpoint and cross section}
\label{sec:3.3.2}
\begin{figure}
\begin{center}
\resizebox{0.4\textwidth}{!}{%
  \includegraphics{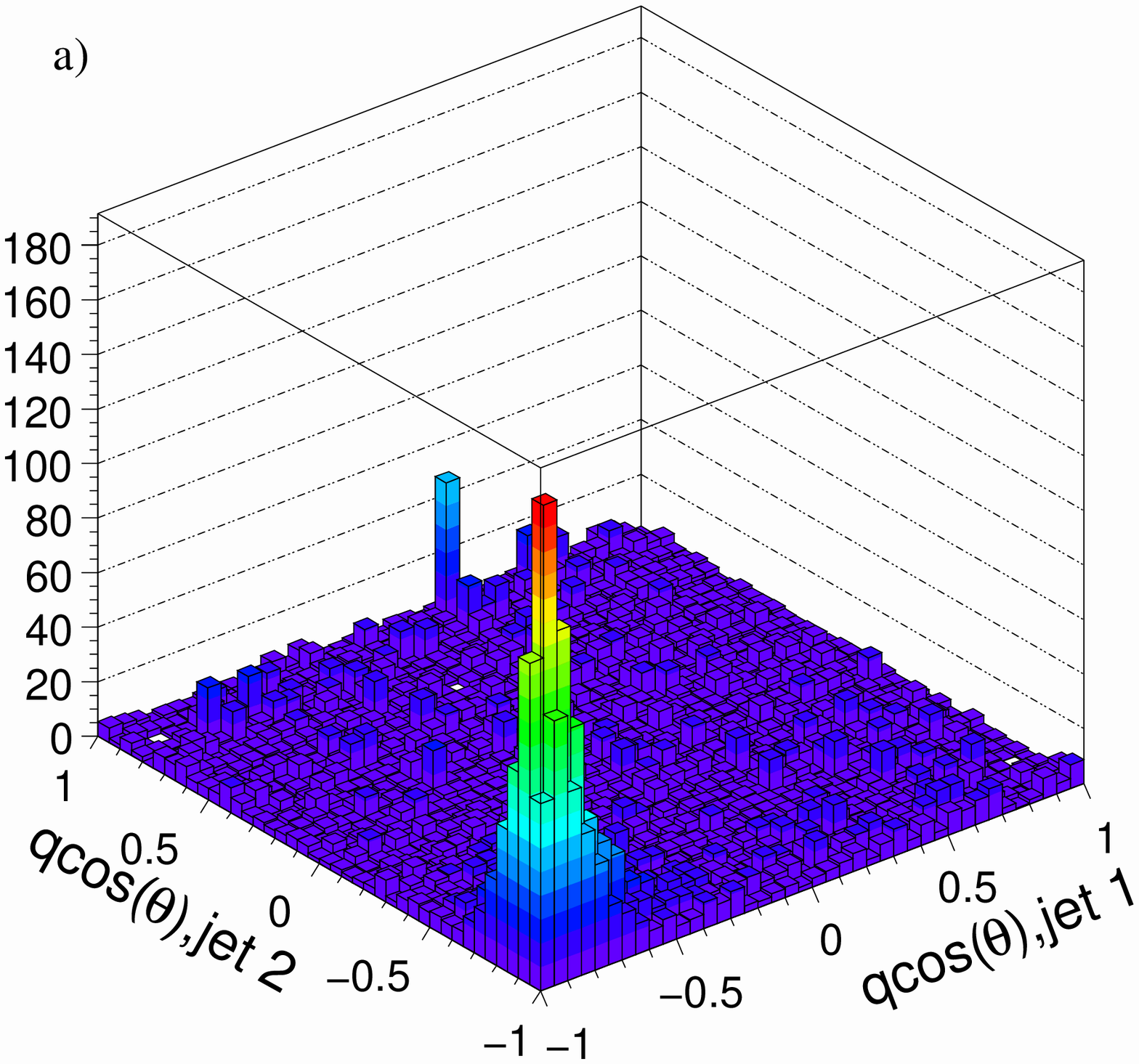}
}
\resizebox{0.4\textwidth}{!}{%
  \includegraphics{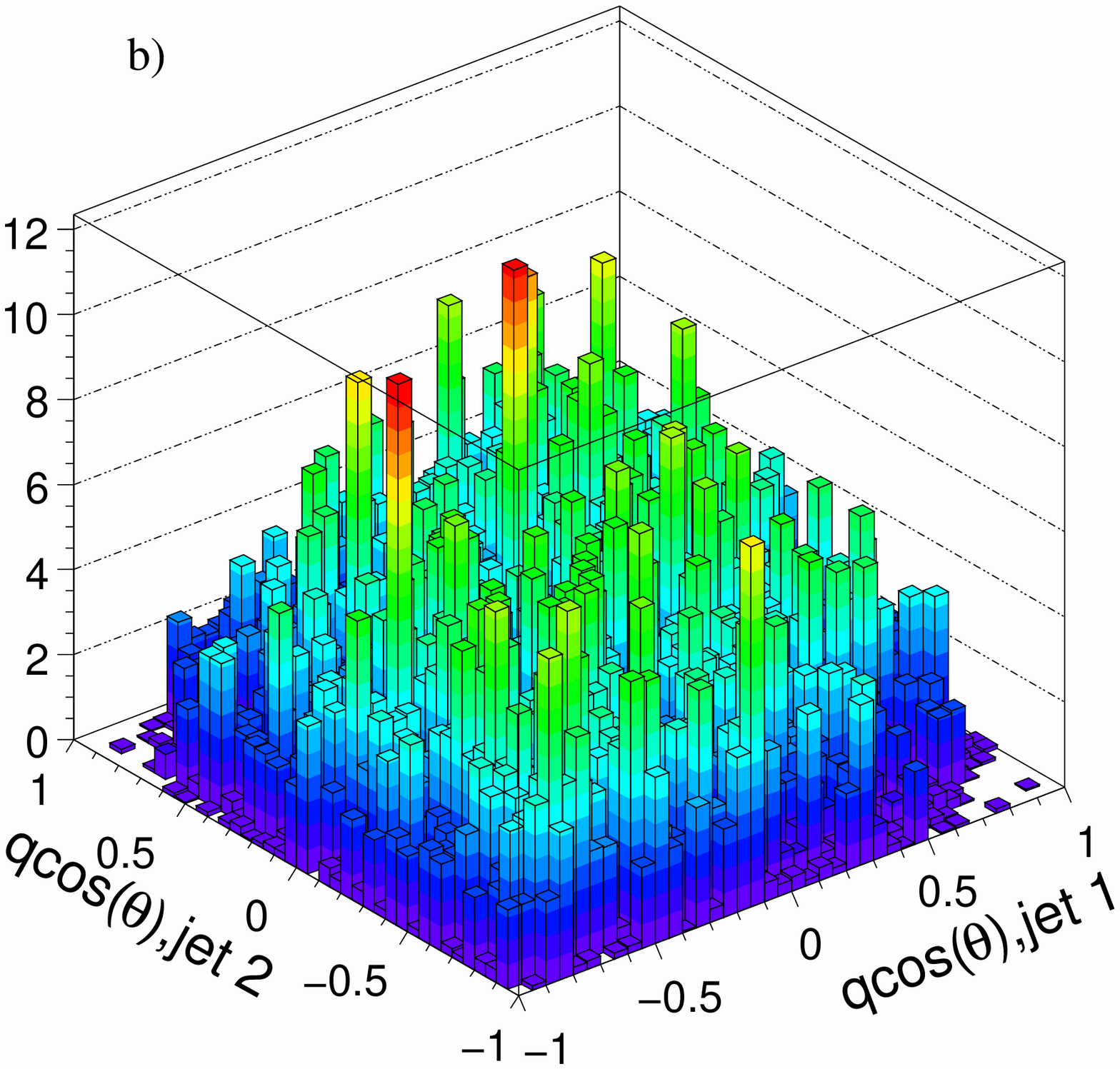}
}
\\
\resizebox{0.4\textwidth}{!}{%
  \includegraphics{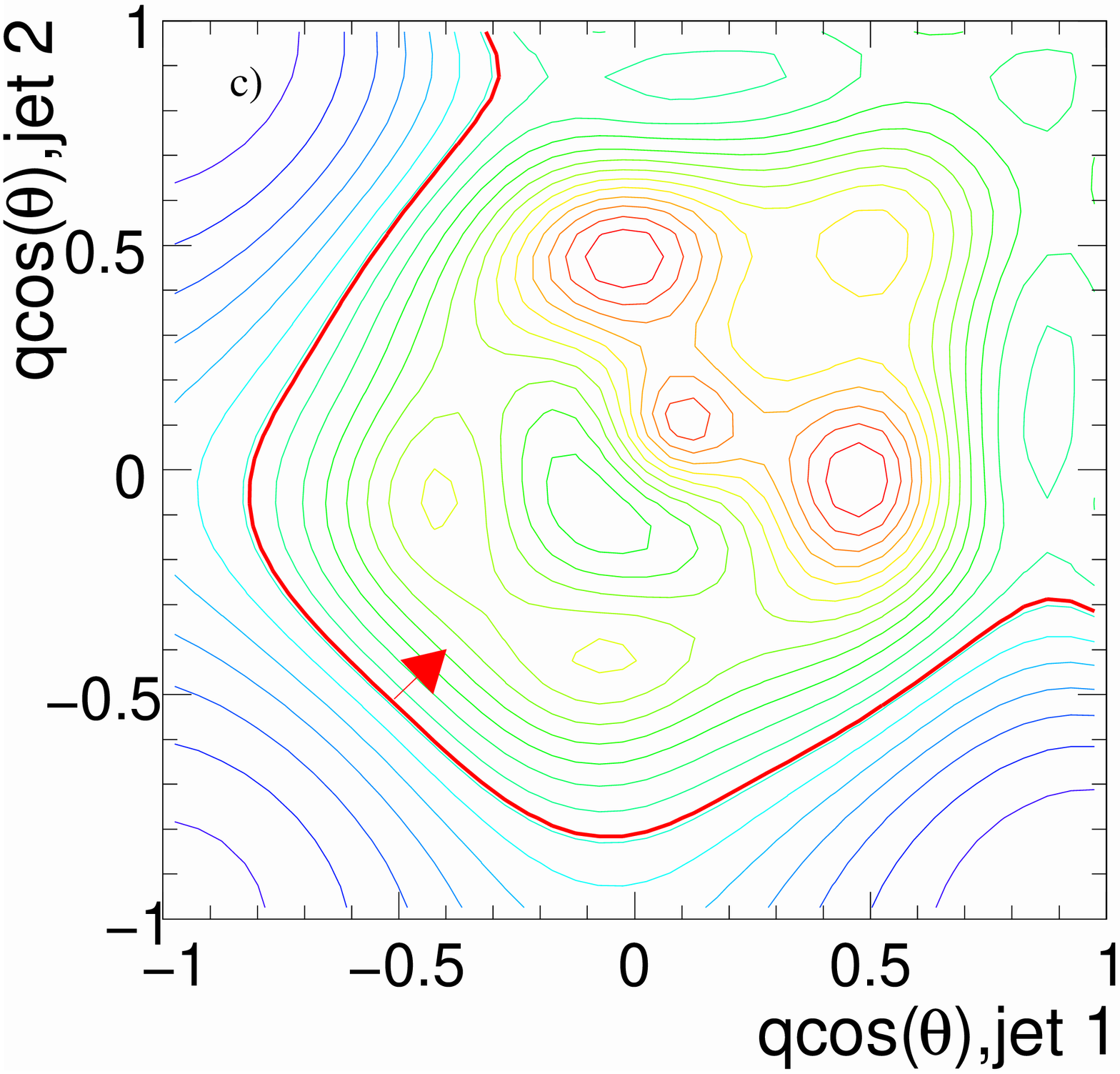}
}
\resizebox{0.4\textwidth}{!}{%
  \includegraphics{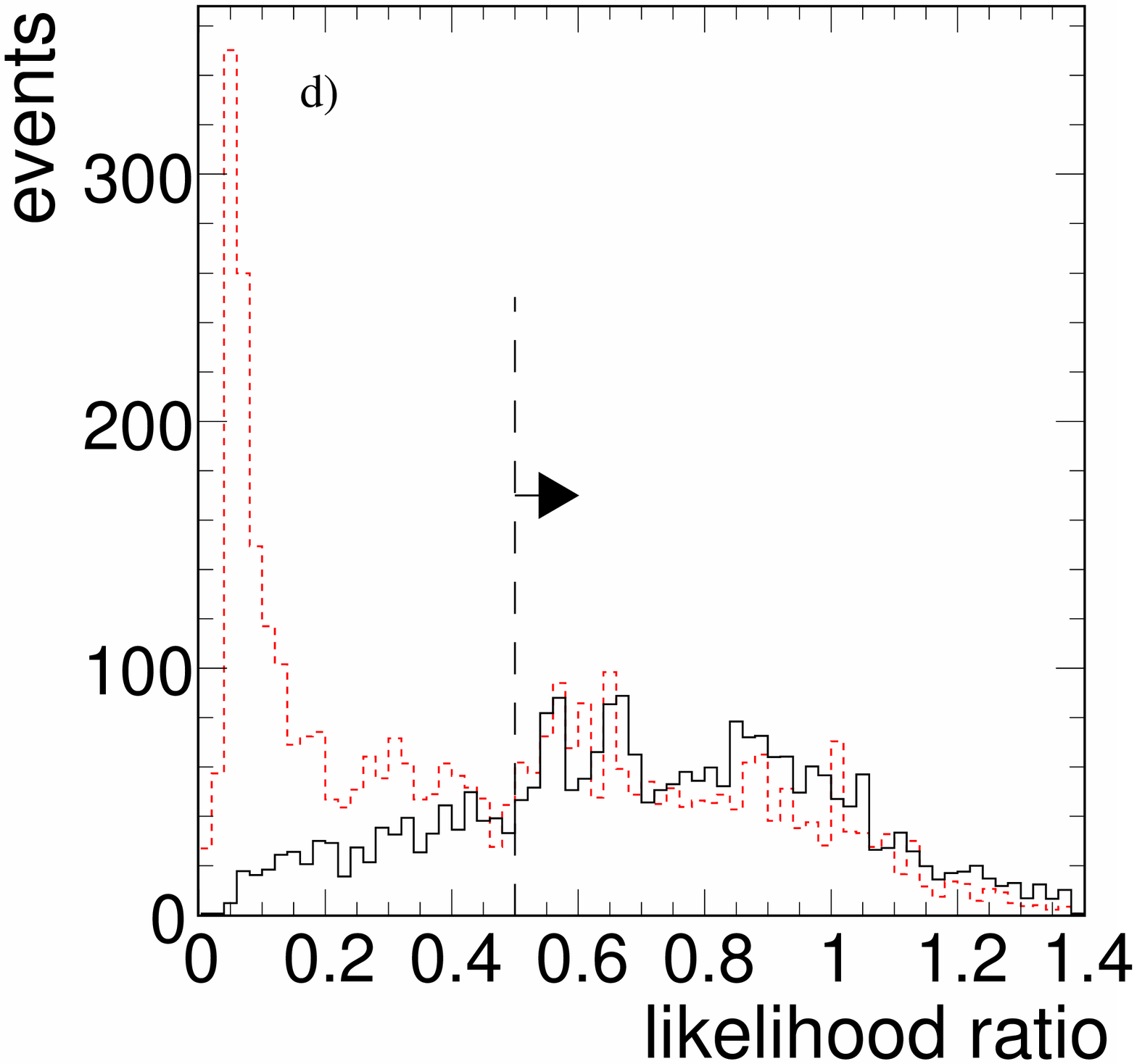}
}
\end{center}
\vspace{-2mm}
\caption{Upper two figures: lego plots of $q_{jet} \cos{\theta_{jet}}$ for the two jets.
a): $WW$ events, b): signal. Lower two figures: c) multi-quadratic fit to
the ratio between signal and background. The thicker line 
indicates the cut: events inside this contour
were accepted, as indicated by the arrow. d): Distribution of the
likelihood ratio for signal (solid black) and background (dashed red).
The vertical line indicates the cut.}
\label{fig:st2_qcos}       
\end{figure}

For the $\sttwo$ sample,
it was demanded that
$E_{vis}>50 ~\GeV$, and
$\Phi_{acop}<155^\circ$, and the energy of the ROE-group was below 10 ~\GeV.
These cuts
left 8095 SM background events, 3500 of which were from
$WW \rightarrow l \nu l \nu$, the rest being $ZZ \rightarrow ll\nu\nu$
or six-fermion events.
In the $\sttwo\sttwo$ sample, 3156 events remained, while
360 SUSY background and 339 $\stone\stone$ events also passed the cuts. 
As the main background at this level was found to be
$WW \rightarrow l \nu l \nu$, it was requested that the
other jet was not a single electron or muon, since the probability, in
$WW \rightarrow l \nu l \nu$ events, 
that a genuine
$\tau$ from one $W$-decay is accompanied by an electron or muon on the
other side is 78~\%, while it is only 35~\% for a $\tau$
from a $\stau$ decay.
At this point, most SUSY background events had
a maximum kinematically allowed jet energy well
below the endpoint of the $\sttwo$ spectrum.
Only $\smul\smul$  
and $\sell\sell$ events where one of the sleptons decayed in a
cascade via a $\XPM{1}$ to a $\tau$ remained. These events were
rejected by requiring that the most energetic jet should not
be an electron or a muon.
Finally, a likelihood ratio was formed of the joint distributions
of $q_{jet1} \cos{\theta_{jet1}}$ and $q_{jet2} \cos{\theta_{jet2}}$
for signal and SM background, as shown in
Fig.~\ref{fig:st2_qcos}. The jet charges $q_{jet,i}$ are determined by the 
charge sum of the tracks in the jet, without any momentum weighting.
The ratio of the two distributions was symmetrised along the diagonal and fitted 
with a multi-quadratic function, in order
to smooth out the statistical fluctuations.
The likelihood ratio exploits the fact that the distribution
of lepton momenta in $WW \rightarrow l \nu l \nu$ events
is forward peaked and forward-backward asymmetric, contrary to the signal.
The final SM background was 2257 jets in 1533 events,
while the signal was 3227 jets in 1972 events (22.3~\% efficiency).
In addition 418 (4) jets in 233 (3) events survived in the $\stone\stone$
(SUSY background) sample.

\begin{figure}
\begin{center}
\resizebox{0.55\columnwidth}{!}{%
  \includegraphics{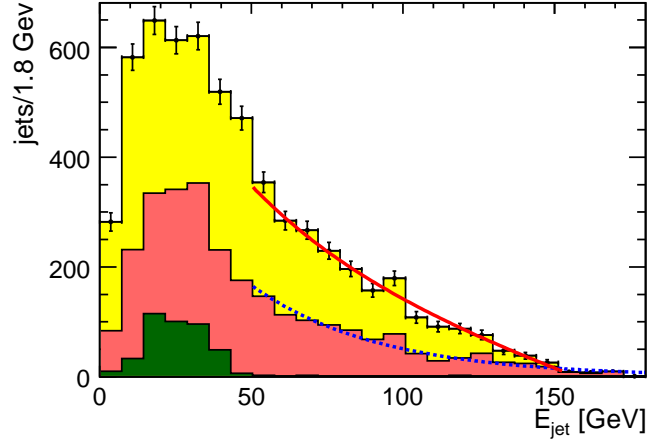}
}
\end{center}
\caption{The jet energy spectrum of events selected in the $\sttwo$ endpoint analysis, with one entry per jet.
Light grey (yellow) histogram: signal, grey (red): SM background, 
dark grey (green): SUSY background. 
The fit to
the background is indicated by the dashed (blue) line. It was fitted to the background only
simulation.
Fit to total sample: solid (red) line. The endpoint is at the cross-over of the
two lines.}
\label{fig:st2mass}       
\end{figure}

The endpoint of the spectrum was determined by first estimating the
background.
As virtually no SUSY background was present at the endpoint,
it was assumed that the expected background level was
known to a much better precision than its Poisson
fluctuations.
The simulated background was parametrised by an
exponential function,  
and the signal was fitted by a straight line added to the
exponential. The spectrum and the fitted functions
are shown in Fig.~\ref{fig:st2mass}.
It should also be pointed out that only the
decay mode $\sttwo \rightarrow \tau \XN{1}$ contributes in
the region where the fit was done: the cascade decays
via a $\XPM{1}$ yield jets of energies of at most 25~$\GeV$.

The endpoint was found to be 
$E_{\tau,max} = 150.2^{+2.0}_{-1.7} ~\GeV$.
In this case, the simple straight line fit is sufficient:
the value found is compatible with
the true value ($152.2 ~\GeV$).
Fitting the mass gave
$\msttwo = 183^{+11}_{-5}~\GeVcc$, assuming the true
value for $\MXN{1}$. At this point, $\msttwo$ is
very sensitive to $\MXN{1}$: $d\msttwo/d\MXN{1}$ = 18,
so an error of 80~$\MeVcc$ on  $\MXN{1}$ translates
into an additional error of 1.4 $\GeVcc$ on $\msttwo$.
The relative error  
$\Delta(\msttwo)/\msttwo$~is 3.6 \%, which is dominated by the
error on the end-point measurement.

\begin{figure}
\begin{center}
\resizebox{0.55\columnwidth}{!}{%
  \includegraphics{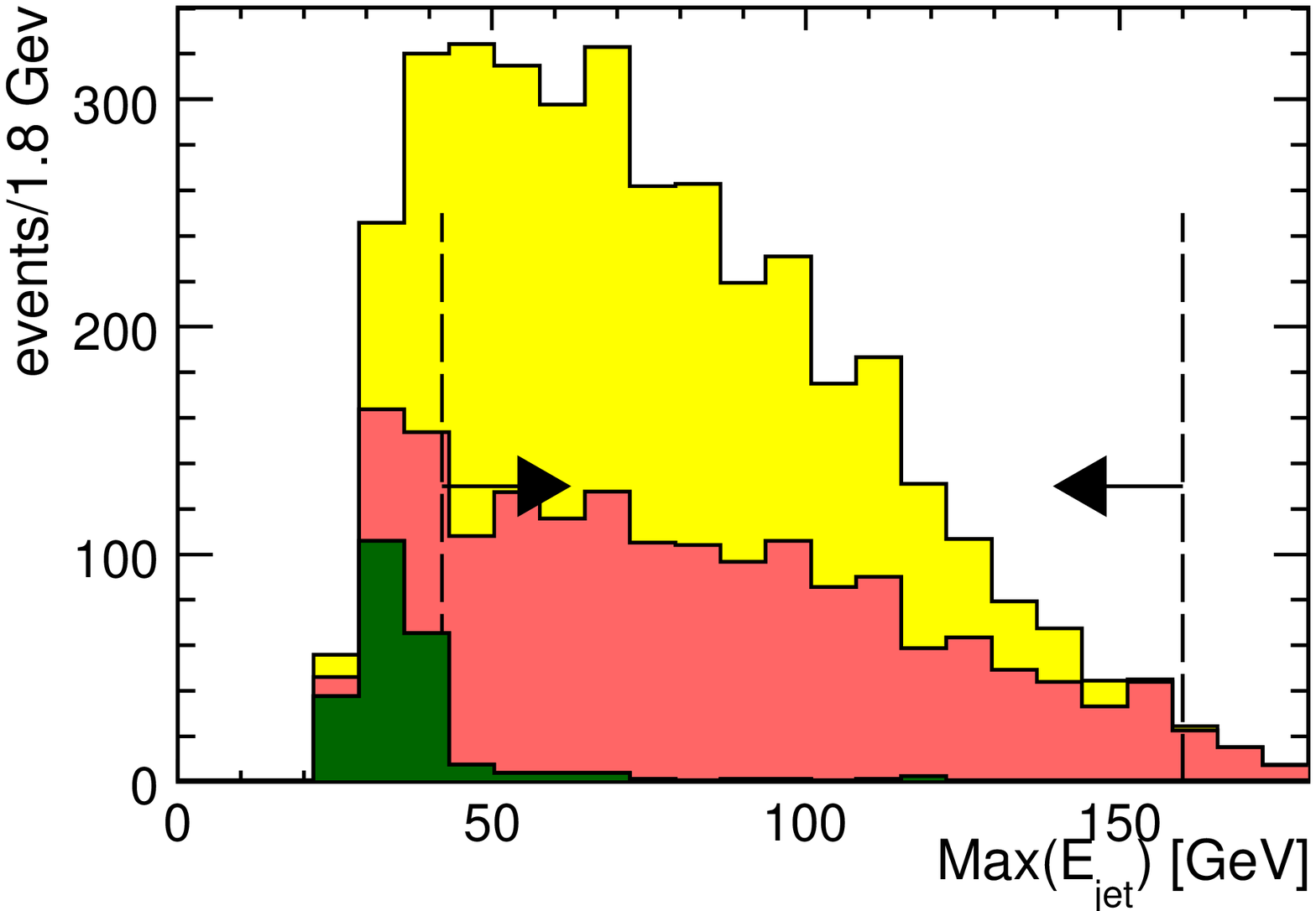}
}
\end{center}
\caption{The spectrum of the maximum jet energy of events selected in the $\sttwo$ 
cross section analysis, with one entry per event.
Light grey (yellow) histogram: signal, grey (red): SM background, 
dark grey (green): SUSY background. 
The cross section was estimated by the number of events having
max($E_{jet}$) between 42 and 160 $\GeV$, as indicated by the vertical lines.}
\label{fig:eps-st2}       
\end{figure}

To determine the ``effective'' cross section,
a sample of events was extracted consisting of those where
any of the two jets were accepted for the endpoint measurement.
The events which had no jet with energy above the endpoint
in $\stone$ production (42 $\GeV$) or with one jet
well beyond the endpoint of $\sttwo$ production (160 $\GeV$) were excluded.
With these additional conditions, the sample contained
1765 $\sttwo$ events, 1374 SM ones, and only 33 other SUSY
events. 
The distribution of events is shown in Fig.~\ref{fig:eps-st2}.
This yielded 
$\Delta(N_{signal})/N_{signal}=4.2~\%$,
assuming again a 100\% uncertainty on the SUSY background, and 
a negligible uncertainty on the SM background.

The corresponding ``effective'' cross section was
$\sigma_{eff}$= $N_{signal}/(\mathcal{L} \epsilon_{signal})$= $17.7\pm 0.7$ fb, 
where once again the uncertainty on $\mathcal{L}$ and $\epsilon_{signal}$
are assumed to be negligible compared to $\Delta(N_{signal})$.

Also for the $\sttwo$, the relative error on the beam energy
is small compared to that of the error on the cross section, so that
$\Delta(\msttwo)/\msttwo$ = $(\Delta(\sigma)/\sigma)(\beta^2)/3(1-\beta^2)$
= 2.4~\%.
The relative error on the mass obtained from the
cross section is thus as small
for $\sttwo$ as for $\stone$ although the cross section is more than
10 times smaller and the background is much larger. This is due
to the fact that there is almost no (poorly known) SUSY background
in the former, and that it is on the rising edge of the cross section
vs. $\sqrt{s}$ curve at $\sqrt{s}$ = 500~$\GeV$.

If, finally, the values of the endpoint and $\msttwo$ were used to
calculate $\MXN{1}$, one finds an error of 1.7~$\GeVcc$, not
including the error on  $\theta_{mix}$.

\subsection{$\tau$ polarisation}
\label{sec:3.4}

In this analysis, the $\tau \rightarrow \pi^{\pm} \nu_{\tau}$
and $\tau \rightarrow \rho^{\pm} \nu_{\tau} \rightarrow \pi^{\pm} \pi^0 \nu_{\tau}$
modes have been studied~\cite{bib:peterthesis}. These modes have already been the subject 
of fast simulation analyses, e.g. in~\cite{bib:diagonal} and~\cite{bib:polarisation}. 

As explained in the introduction, 
the compositions of the $\stau$ and
the $\XN{1}$ manifest themselves in the probability that
the $\tau$ has either helicity, i.e. in the $\tau$ 
polarisation, $\mathcal{P}_\tau$.
To observe $\mathcal{P}_\tau$, one uses the fact that only one helicity
state exists for $\nu_{\tau}$, which induces a angular distribution
$\propto 1-a \mathcal{P}_\tau \cos{\psi}$ for the visible decay products in
the $\tau$ rest frame~\cite{bib:taupolbase} ($ \psi$ is the angle between the 
helicity axis and the momentum, and $a$ is a factor
depending on the mass and spin of the $\tau$ decay product).
The lab frame energy spectrum
is therefore sensitive to $\mathcal{P}_\tau$.
If, in addition, the decay product is a vector, the probability of
the product being transversely or longitudinally polarised also 
depends on $\mathcal{P}_\tau$: one case would correspond to a 
(more likely) conservation of
helicity, the other to a (less likely) helicity flip.
Whether the vector particle is transverse or longitudinal can be
determined by the angular distribution of its decay products:
in the rest frame of a two-body decay to scalars, the products 
tend to line up along the spin axis in the longitudinal case ($\xi=0$),
and to be perpendicular to it in the transversal case ($\xi=\pi/2$).
In the lab frame, the longitudinal case therefore corresponds to
a case where the energies of the two scalars are maximally different,
while in the transversal case they tend to be quite similar.
In particular, the variable $R=E_1/(E_1+E_2)$ is distributed
as $(1-\beta_{vector}\cos(\xi))$, which is quite insensitive
to the lab frame energy of the vector particle, once it is substantially
larger than its mass (i.e. $\beta \approx 1$).
$R$ will be distributed as $(R-1/2)^2$ for the 
longitudinal case, and as $1/4-(R-1/2)^2$ in the 
transversal case~\cite{bib:diagonal}~\cite{bib:taupolbase}.
In SPS1a', the mixing is not particularly large\footnote{Due to a somewhat
unlucky choice of conventions, this statement corresponds to
$\theta_{mix}$ close to $\pi/2$: The left-handed field is considered
as the first chiral state, while the lighter state is considered
as the first mass state. However, in the mass matrix, 
the diagonal element corresponding to the right-handed
state is expected to be the smaller, so the lighter state in the
case of the off-diagonal elements being zero is the pure right-handed
one. Hence, with the convention, the transformation chiral state $\rightarrow$
mass state turns the labelling ``up-side down'', hence a small
mixing corresponds
to $\theta_{mix}$ close to $\pi/2$.}, and $\stone$ is expected
to be mainly right-handed. Hence,  $\mathcal{P}_\tau$ is expected
to be rather close to +1, and the spectrum in the $\tau \rightarrow ~ scalar$
should be harder than for the other helicity. In the case  $\tau \rightarrow ~ vector$,
the vector meson is mainly longitudinal, yielding an $R$ distribution
peaking close to 0 and 1.

\subsubsection{The $\tau \rightarrow \pi^{\pm} \nu_{\tau}$ channel}
\label{sec:3.4.1}
\begin{figure}[hb]
\begin{center}
\resizebox{0.55\columnwidth}{!}{%
  \includegraphics{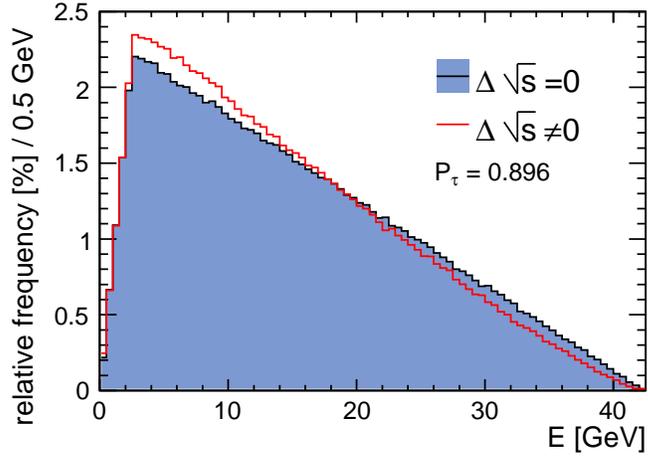}
}
\end{center}
\caption{The simulated energy spectrum the pions in 
$\stone  \rightarrow \tau \rightarrow \pi^{\pm} \nu_{\tau}$. The filled histogram
shows the distribution for $E_{CMS} \equiv$ 500 $\GeV$ and ISR switched
off in the generator, while the open one shows the spectrum with
the ILC beam spectrum and ISR included.}
\label{fig:truepispect}       
\end{figure}

The spectrum of the pions in the decay chain $\stau \rightarrow \tau \XN{1}
\rightarrow \pi^{\pm} \nu_{\tau} \XN{1}$ is shown in Fig.~\ref{fig:truepispect},
with and without ISR and beam energy spread.
As the effect of these two factors clearly are not negligible,
the true spectra were determined for extreme polarisations
($+1$ or $-1$), and parametrised correction functions
were calculated for both cases. These functions were
double polynomials of degree 2,
the two pieces being applied above or below $E_{\tau , min}$, respectively.
With these parametrisations ($F(E,+1)$ and $F(E,-1)$) at hand,
the true spectrum for any polarisation can be obtained by
applying the combined correction
$$
F(E,\mathcal{P}_\tau) = \frac{ 1 + \mathcal{P}_\tau }{ 2 } 
F(E,+1) + \frac{ 1 - \mathcal{P}_\tau }{ 2 }  F(E,-1) 
$$ 
It should be noted that the highest sensitivity to the polarisation is
in the region with  $E_\pi < E_{\tau,min}$.

To extract the signal, the cuts described in Sects. \ref{sec:3.1}
and \ref{sec:3.2}
were supplemented by demanding that $E_{vis}$ be less than $90~\GeV$,
and that none of the jets had an energy exceeding $60~\GeV$.
The events should contain at least one signal decay candidate,
defined as a jet that  only contained a single particle, and that that particle
was charged.
After these cuts,  there were 8.41$\times 10^4$ SM events
(839 non $\gamma\gamma$), 201 $\sttwo\sttwo$ events, and 
1678 other SUSY events remaining. 10730 $\stone\stone$ events remained,
i.e. 21460 $\stone$ decays. In 4047 of these, the subsequent $\tau$-decay
was the signal-channel $\tau \rightarrow \pi \nu_{\tau}$.

To reject  $\tau \rightarrow \ell \nu_{\tau} \nu_{\ell}$ and
$\tau \rightarrow K \nu_{\tau} $ from the sample of signal candidate jets,
the full power of
particle identification of the ILD was employed: The result from the
PFOid package, which is based on the calorimetric
measurements, was supplemented by the measurement of $dE/dx$ in the TPC.
Only about 0.4~\% of the non-signal decays were misidentified,
while the efficiency to accept signal decays was 80~\%. 

This requirement was also very efficient in rejecting the remaining
$\gamma\gamma$ background, since only a small fraction of these events
did contain two $\tau$ leptons. The same was true for 4- and 6-fermion
background, albeit to a lesser extent. The background from non-$\stone\stone$
SUSY channels, on the other hand, largely contained two $\tau$ leptons and was
reduced only sightly more than the signal. It was nevertheless 
concluded that no
further cuts were needed, and
the final selection contained 3311 signal jets, 126 other decay modes
of $\stone$, 334 other SUSY decays, and 122 SM jets.

The procedure to extract the polarisation in the presence of background
was to first fit  a heuristic function to the simulated background 
alone\footnote{When real data is available, the simulation of the background
can be verified by reversing cuts to select a signal-free, but
SUSY-dominated region in the parameter space.}.
The signal
selection cuts were then applied to the signal+background
sample,
and the function describing the background 
was subtracted from the observed distribution.
An efficiency correction function, determined from signal-only
simulation, was applied. As the efficiency could possibly be
dependent on the helicities of the two $\tau$ leptons in the event, 
the efficiency correction was
para\-metrised as
\begin{align*}
\epsilon(E,\mathcal{P}_\tau) = &\left ( \frac{ 1 - \mathcal{P}_\tau}{ 2 } \right )^2
\epsilon_{--}(E) +  \left ( \frac{ 1 - \mathcal{P}^2_\tau }{ 2 } \right ) \epsilon_{+-}(E)+ \\
 &\left ( \frac{ 1 + \mathcal{P}_\tau }{ 2 } \right )^2 \epsilon_{++}(E)
\end{align*}
The efficiencies $\epsilon_{--}(E),
\epsilon_{+-}(E)$, and $\epsilon_{++}(E)$, correspond to the cases
of the $\tau$ leptons being both of negative, of opposite, or both of
positive helicity, respectively. These functions were
separately determined from dedicated fast simulation samples with
the corresponding helicity configurations.

\begin{figure}[hb]
\begin{center}
\resizebox{0.55\columnwidth}{!}{%
  \includegraphics{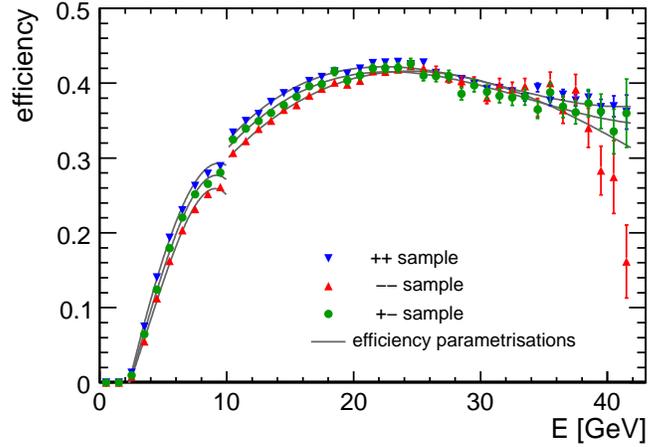}
}
\end{center}
\caption{Ratio of the spectra before and after selection (points),
and fitted efficiencies (lines) for different combinations of $\tau$ helicity.
Inverted triangles (blue): both $\tau$ leptons are
right-handed. Circles (green): the $\tau$ leptons have opposite helicity.
Triangles (red): both $\tau$ leptons are left-handed. The discontinuity 
at 10~$\GeV$ is due to the fact
that the particle identification algorithm changes 
from one set of p.d.f.s to another at that energy.}
\label{fig:epspol}       
\end{figure}

The ratios between initial and selected spectra are
shown in Fig.~\ref{fig:epspol}, together with the fitted
efficiency functions. 
A slight dependence on $\mathcal{P}_\tau$ was indeed
observed
and was found to be primarily caused by the cut on $\rho_\perp$.

\begin{figure}[hb]
\begin{center}
\resizebox{0.55\columnwidth}{!}{%
  \includegraphics{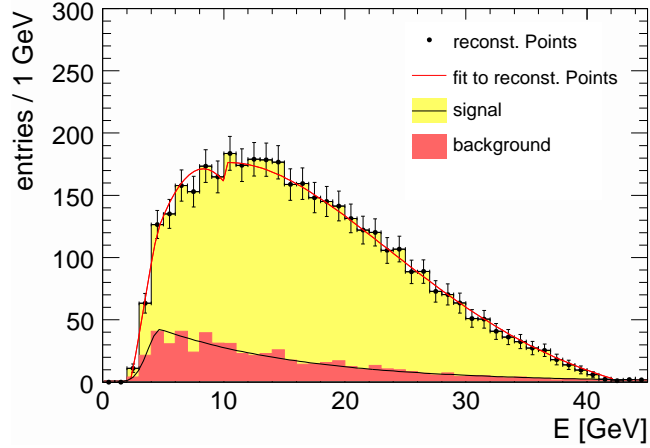}
}
\end{center}
\caption{The energy spectrum of the selected pions. Light grey (yellow) 
histogram: signal. Dark grey (red) histogram: background.
Black line: background fit. Grey (red) line: overall fit.
The ``dent'' in the overall fit at $E_{\pi}$=10 $\GeV$ is due to the 
discontinuity of the efficiency parametrisation, see Fig.~\ref{fig:epspol}.}
\label{fig:pispect}       
\end{figure}

\begin{figure}[hb]
\begin{center}
\resizebox{0.55\columnwidth}{!}{%
  \includegraphics{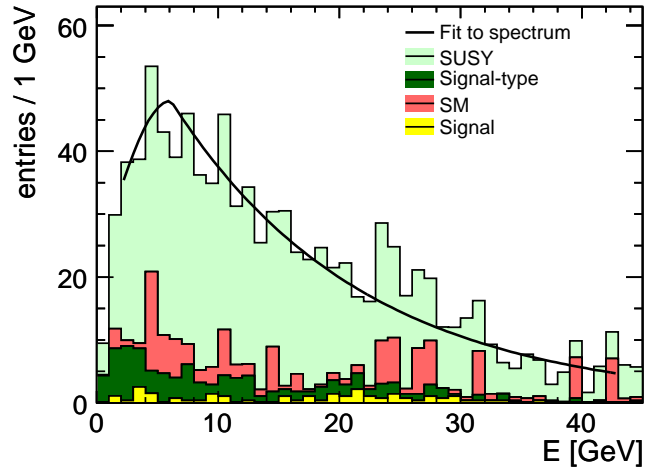}
}
\end{center}
\caption{Energy spectrum of the pion candidates in the control sample.
Light grey (light green) histogram: SUSY background (non-$\stone$). 
Grey (red) histogram: SM. Dark grey (dark green) histogram:
SUSY background from other decays 
of the $\tau$ from $\stone$. Medium light grey (yellow) histogram:
signal. Line: fit to spectrum.}
\label{fig:pol_control}       
\end{figure}

The resulting distribution was then fitted
with the theoretical spectrum, corrected for ISR and beam spread,
and the polarisation was obtained, see  Fig.~\ref{fig:pispect}.
Assuming an integrated luminosity of 500 fb$^{-1}$,
the value found was  $\mathcal{P}_\tau = (91 \pm 10)$~\%,
where the error is statistical.
The expected value in SPS1a' is 89.6~\%.
The fitted normalisation and the
polarisation showed a quite sizable correlation, so if the 
normalisation was
calculated using the value and uncertainty 
of the ``effective'' cross section
obtained in section ~\ref{sec:3.3.1},
the error on the polarisation decreased to 6~\%.
The uncertainty of the average background stemmed
from the uncertainty on its SUSY component,
while the average of the SM component would have a negligible
uncertainty. 
No signal-free sample with composition and spectrum
close to that of the background in the selected sample could
be constructed in this ana\-lysis.
Hence, the background cannot be determined from the data
itself, and one must
rely on MC modelling.
One could assume that no other data set than
the ILC data will exist to validate a SUSY simulation, so
the uncertainty of the model would be determined by
the uncertainty of the ILC data itself.
It is also essential that the sample used to verify the
the SUSY simulation contains as little of the signal-channel
as possible.
Such a signal-free sample
was obtained by reversing the
cuts on the invariant mass of the other jet,
the acoplanarity angle, and the cut on 
$(E_{jet 1}+E_{jet 2})\sin{\Phi_{acop}}$. This sample,
shown in Fig.~\ref{fig:pol_control}, contained 829 SUSY
background jets, 128 SM jets, and 26 signal jets.

The derivative of the fitted polarisation w.r.t. variations
in the estimated SUSY background was determined numerically,
and when multiplied by the statistical error on the
determination of the Poisson parameter, it yielded an
additional error on the polarisation of 5~\%.

The influence of  $\mstone$ and $\MXN{1}$ was determined
numerically, by separately varying $\MXN{1}$ and $E_{\tau,max}$
in the fits, as the measurement of these two quantities
are largely independent. A close to linear dependence
on both these variables was found, and by using the
uncertainty on $E_{\tau,max}$ from Sect. ~\ref{sec:3.3.1},
and assuming  $\sigma_{\MXN{1}} \approx 250~\MeVcc$,
an additional uncertainty of 3.4~\% was determined.

Hence, the final result was
$$
\mathcal{P}_{\tau}=93 \pm 6 \pm 5 ~\mathrm{(bkg)} \pm 3 ~\mathrm{(SUSY masses)}~\%
$$

\subsubsection{The $\tau \rightarrow \rho^{\pm} \nu_{\tau} \rightarrow \pi^{\pm} \pi^0 \nu_{\tau}$ channel}
\label{sec:3.4.2}
In the $\rho$ channel, the observable sensitive to the
polarisation is $E_{\pi}/E_{jet}$ which - as  mentioned above - is 
expected to be
insensitive to the exact value of $E_{jet}$, and hence to beam spectrum and 
ISR effects. Therefore,
no re-evaluation of the true spectrum due to these
effects is needed.

Also in the $\rho$ channel, the cuts described in Sects. ~\ref{sec:3.1} 
and ~\ref{sec:3.2}
were used. 
In addition, it was demanded that  $E_{vis}$ be less than $90 ~\GeV$,
and that none of the jets had an energy exceeding $43 ~\GeV$.
The signal decay candidates, of which there should be at least one in the
event, were selected by demanding that the
jet only contained a single charged particle,
and that it was accompanied by at least two neutral particles.
To further reduce
the  $\gamma\gamma$ background, the cut on $\rho_\perp$ was tightened to
$\rho_\perp > (3.5\sin{\Phi_{acop}}+2)~\GeVc$.  
After these cuts,  there were 2.93$\times 10^5$ SM events
(733 non $\gamma\gamma$), 736 $\sttwo\sttwo$ events, and 
1373 other SUSY events remaining. 10451 $\stone\stone$ events remained,
i.e. 20902 $\stone$ decays. In 11120 of these, the following $\tau$-decay was 
$\tau \rightarrow \rho^{\pm} \nu_{\tau}$.

\begin{figure}
\begin{center}
\resizebox{0.45\textwidth}{!}{%
  \includegraphics{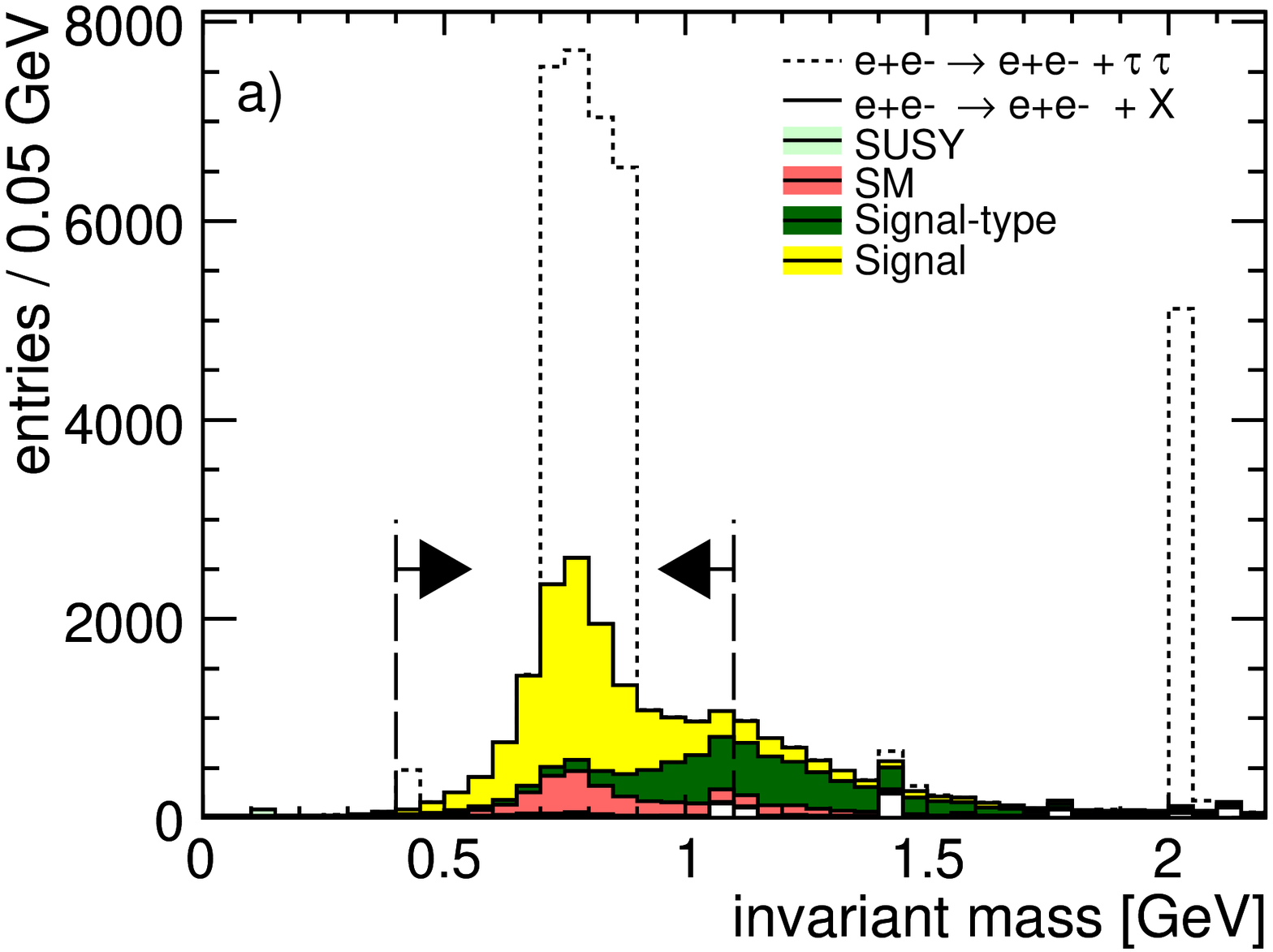}
}
\resizebox{0.45\textwidth}{!}{%
  \includegraphics{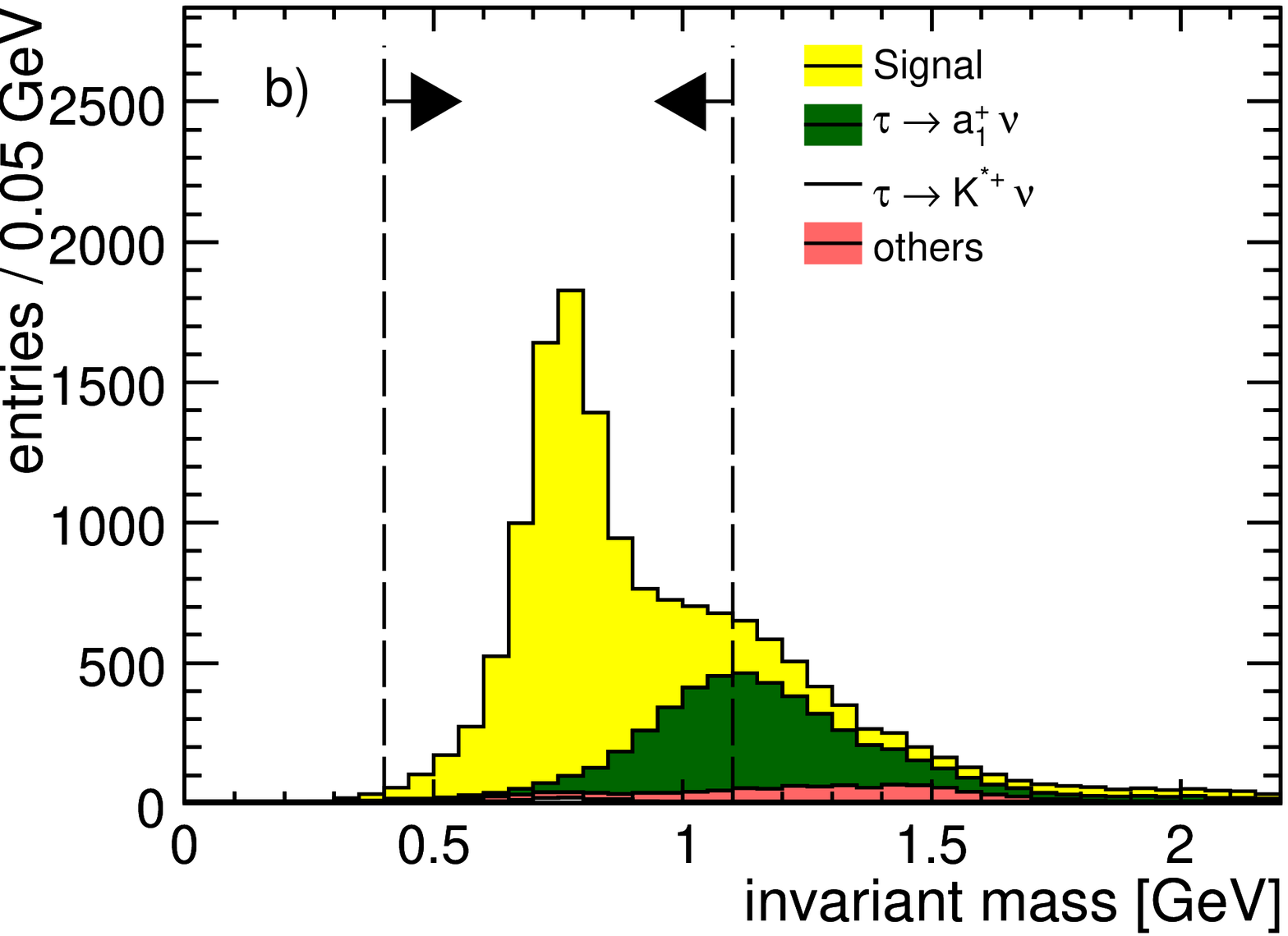}
}
\end{center}
\caption{Distribution of $M_{jet}$ in the selected $\rho$ sample.
a): Signal and all background:
Medium light grey (yellow) histogram: Signal. Dark grey (dark green) histogram:
Background from other decays of the $\tau$  from $\stone$.
Light grey (light green) histogram: Background from other SUSY
processes. Grey (red) histogram: Background from SM processes,
except $\gamma\gamma$. Open dashed and solid histograms:
Background from $\gamma\gamma \rightarrow \tau\tau$, and
$\gamma\gamma \rightarrow X$.
b): Signal and signal-type SUSY background,
broken down by $\tau$ decay-mode.:
Light grey (yellow) histogram: Signal.
Dark grey (green) histogram:
Background from $\tau \rightarrow a_1 \nu_{\tau}$.
Open histogram: Background from $\tau \rightarrow K \nu_{\tau}$.
Grey (red) histogram:
Background from other $\tau$ decays.}
\label{fig:rhomass}       
\end{figure}

The signal decays were selected by demanding that the
corresponding jet had  $|\cos{\theta_{jet}}| < 0.8$.
The calorimeter-based PFOid algorithm was not used,
because of the presence of two or more neutral clusters
close to the track gives an
unacceptably low efficiency for the signal.
The measurement of $dE/dx$ in the TPC has no such problem,
and was used 
to reject  $\tau^\pm \rightarrow e^\pm \nu_{\tau} \nu_{e}$ (accompanied
by bremsstrahlung photons) and
$\tau^\pm \rightarrow K^\pm \pi^0 \nu_{\tau} $. 
Finally, the mass of the jet should be around the mass
of the $\rho$: $M_{jet} \in [0.4,1.1] ~\GeVcc$.
Only about 7~\% of the non-signal decays of the $\stone$ were 
misidentified,
while 86~\% of the signal decays still remained. 
Figure ~\ref{fig:rhomass} shows the
invariant mass spectrum of the selected events.

The final selection contained 
8165 
signal jets, 1991 from other decay modes
of the $\tau$ in $\stone\stone$-events 
($a_1$: 1602, $K^{*\pm}$: 131, all other 258), 
1825 from jets in other SUSY channels, and 195 SM jets.
In addition, the background from $\gamma\gamma$ processes was estimated to be 
3000 jets, but the lack of statistics in the simulation
made it difficult to asses this number with precision.
Due to this, the background was estimated in a somewhat less
sophisticated manner than for the $\pi$ channel. The distribution
of $E_{\pi}/E_{jet}$ for the  $\gamma\gamma$ before cuts was
scaled down to correspond to the number of such events that
survived all cuts, and this rescaled distribution was
added to the background from other sources.

\begin{figure}
\begin{center}
\resizebox{0.31\textwidth}{!}{%
  \includegraphics{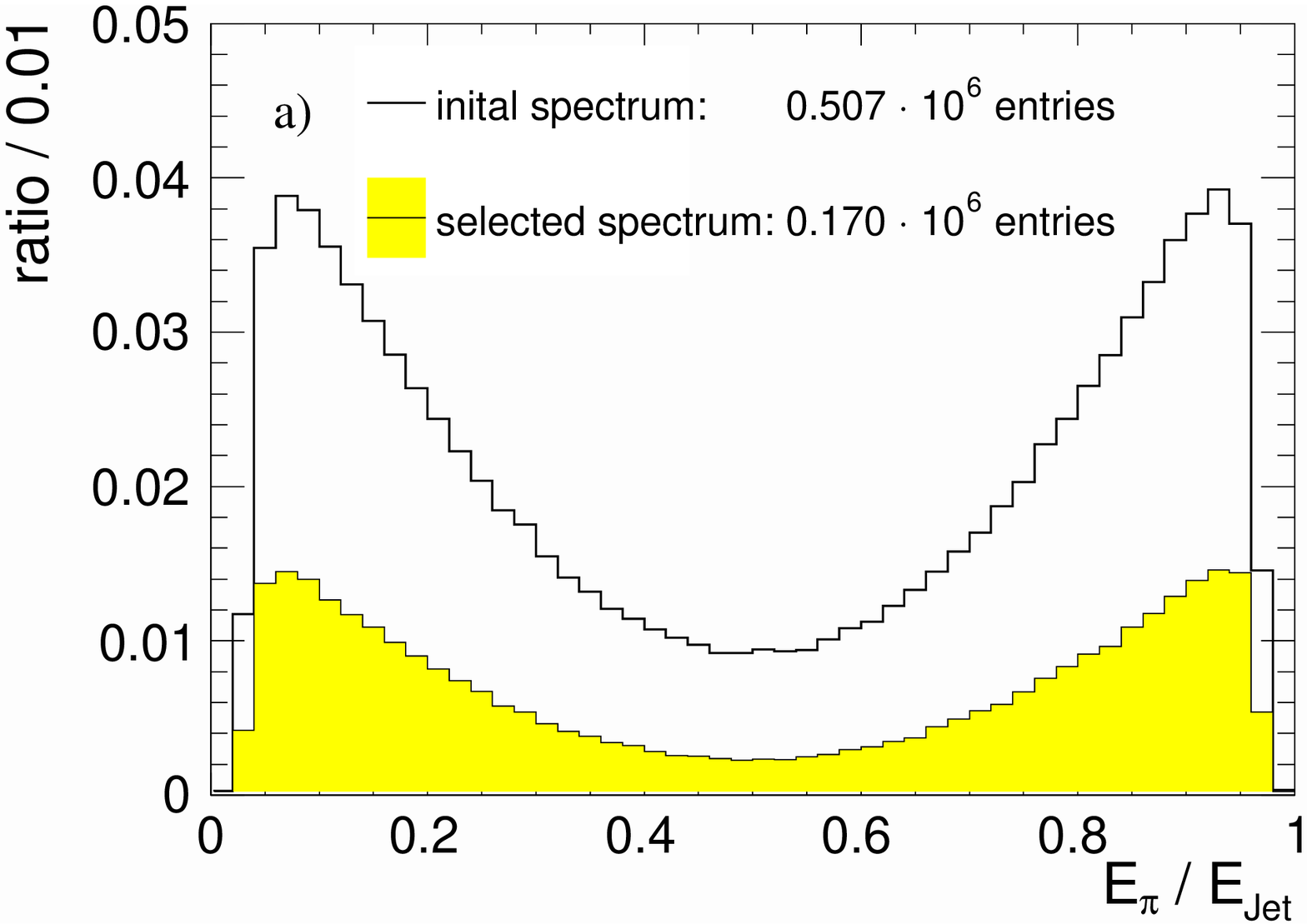}
}
\resizebox{0.31\textwidth}{!}{%
  \includegraphics{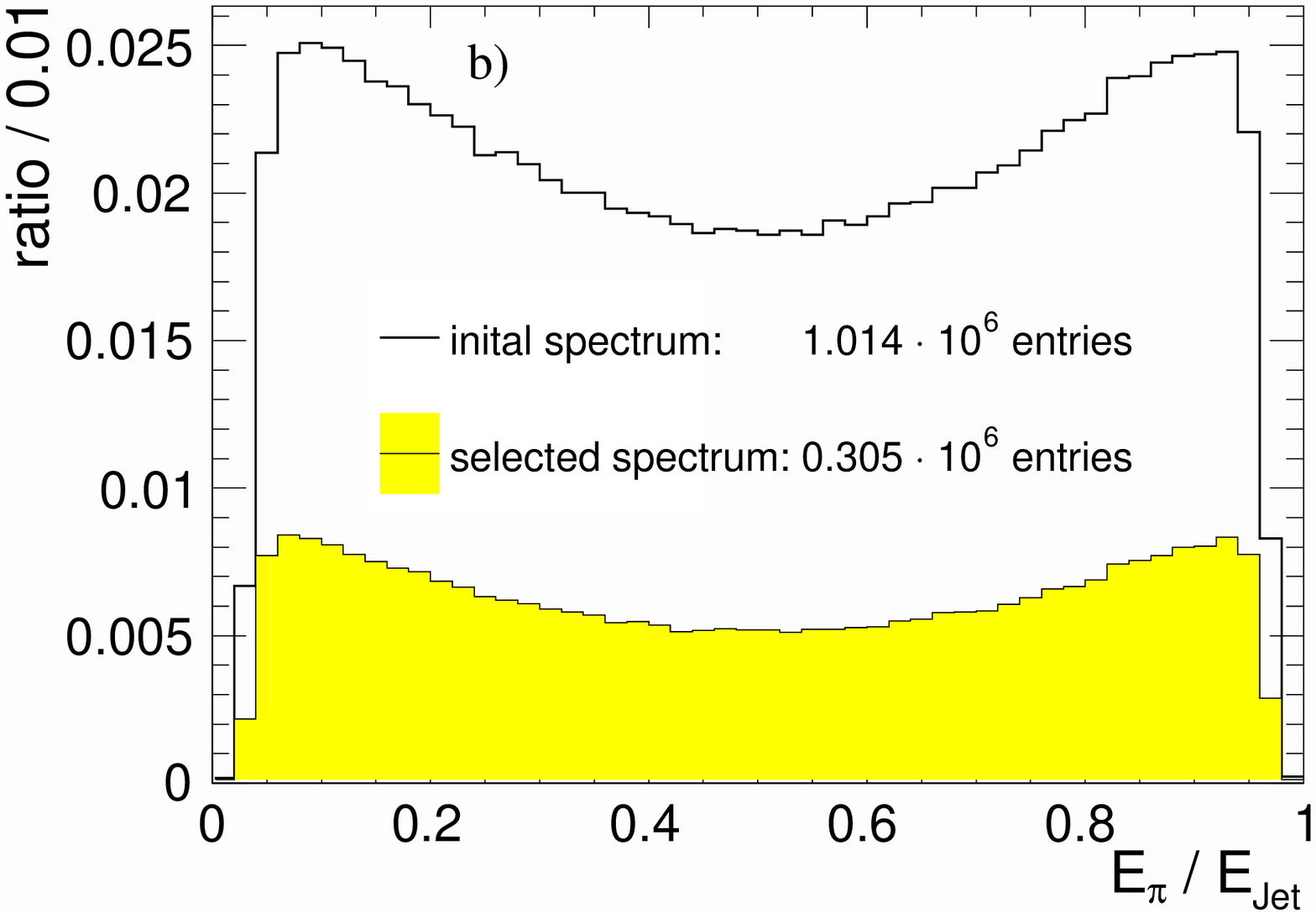}
}
\resizebox{0.31\textwidth}{!}{%
  \includegraphics{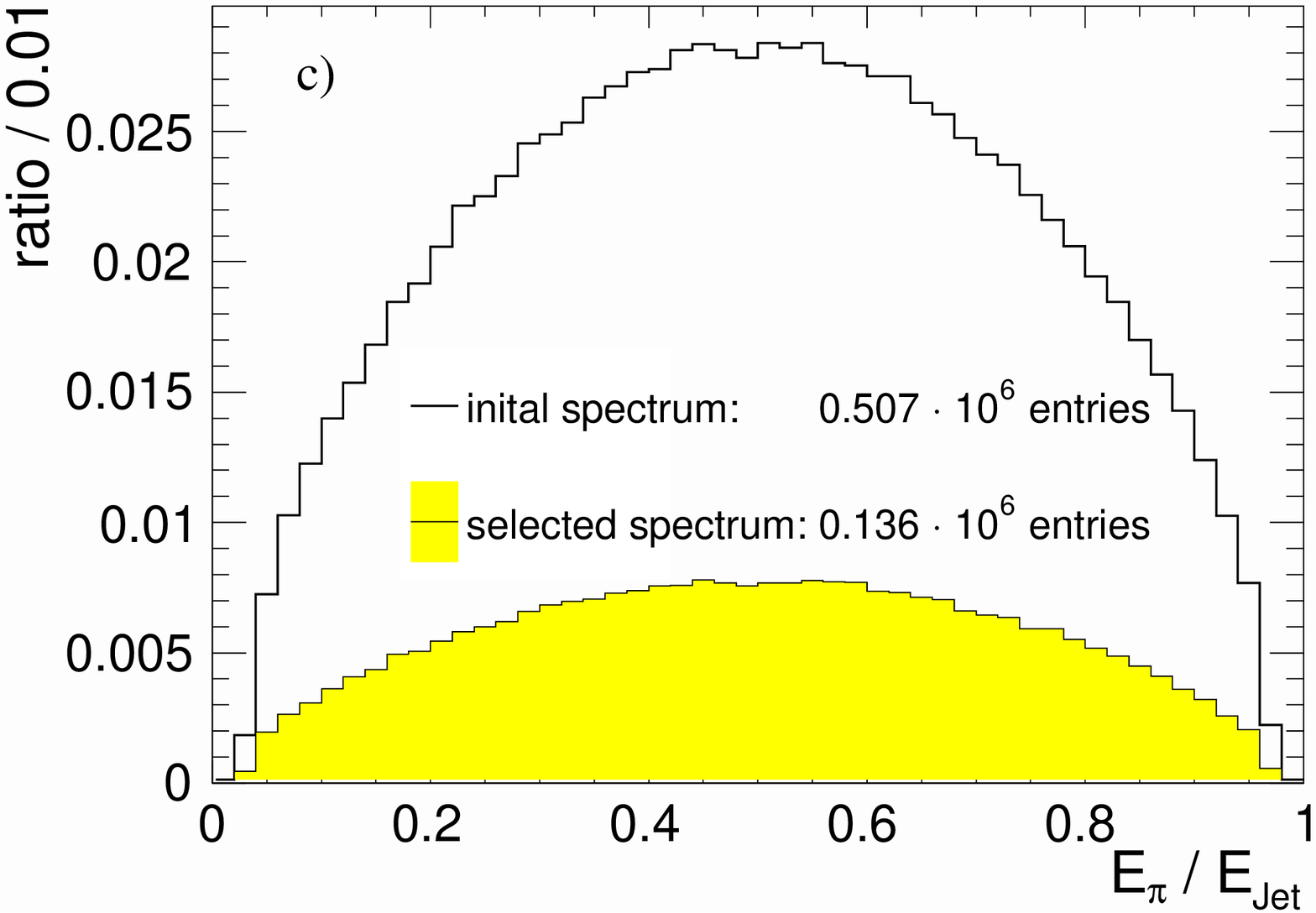}
}
\end{center}
\caption{Distribution of $R=E_{\pi}/E_{jet}$ before (open histogram)
and after event selection (grey (yellow) histogram). a): both $\tau$ leptons are
right-handed. c): the $\tau$ leptons have opposite helicity.
r): both $\tau$ leptons are left-handed.}
\label{fig:specspolrho}       
\end{figure}

Similarly to the $\pi$ channel, an efficiency correction function, determined from signal-only
simulation, was applied. 
Only the efficiency is assumed be
dependent on the experimental situation, not the true spectrum.
Therefore, and contrary to the
case of the $\pi$ channel,
the two steps (spectrum correction and efficiency determination) could
be merged into one, directly yielding an efficiency-corrected
model prediction:
\begin{align*}
dN/dR = N & \left [  \left (  \frac{ 1 - \mathcal{P}_\tau }{ 2 } \right )^2
f_{--}(R) +  \left ( \frac{ 1 - \mathcal{P}^2_\tau }{ 2 } \right ) f_{+-}(R)+ \right . \\
& \left .
   \left ( \frac{ 1 + \mathcal{P}_\tau }{ 2 } \right )^2 f_{++}(R) \right ]
\end{align*}
The efficiency corrected spectra $f_{--}(R)$, $f_{+-}(R)$, 
and 
$f_{++}(R)$, correspond to the cases of
the $\tau$ leptons being both of negative, of opposite, or both
of positive helicity, respectively.
These spectra were determined by fast 
simulation,  see Fig.~\ref{fig:specspolrho}.

\begin{figure}
\begin{center}
\resizebox{0.55\columnwidth}{!}{%
  \includegraphics{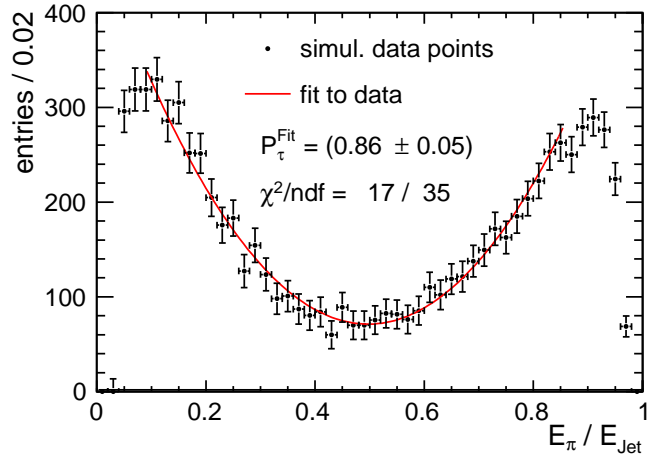}
}
\end{center}
\caption{The distribution of $R=E_{\pi}/E_{jet}$ in the selected sample.
The line shows the fitted efficiency and background corrected
model.}
\label{fig:rhospect}       
\end{figure}

The fast simulation was too optimistic, both in overall selection efficiency,
and the efficiency for low and high $R$. Hence the fit was restricted to 
$R$ between 0.1 and 0.85, where the shape between full and fast simulation
agreed, and the efficiency was scaled down equally for all polarisation
configurations so that it agreed with the full simulation value.
The observed spectrum was then fitted, with $N$ and $P_{\tau}$ as parameters,
see Fig.~\ref{fig:rhospect}.
The result for the polarisation, simultaneously fitted
with the normalisation, was found to be
$\mathcal{P}_{\tau}$ = $86 \pm 5~\%$.
Due to the large uncertainty on the $\gamma\gamma$ contribution, it was
of little use to study the effects of the uncertainty of the
much smaller SUSY background.
Due to the near invariance of the $R$-distribution, the
actual values of $\mstone$ and $\MXN{1}$ are expected to have only a small 
impact on the results.

\section{Summary and Conclusions}
\label{sec:4}
A study of $\stau$ channels in the SPS1a' SUSY scenario based on a full simulation
of the ILD detector at the ILC was presented.

The study was performed in the context of the detector performance
studies in view of the ILD Letter of Intent. It
was therefore based on a full detector simulation of all known SM
processes and machine related backgrounds. 
All accessible channels of the SPS1a' SUSY model
were also simulated with the same procedures.
The simulation  was done
assuming that the ILC was run at a centre-of-mass energy
of 500~$\GeV$, delivering an integrated luminosity of 500~fb$^{-1}$
with the electron beam being 80\% right polarised, and the positron
beam being 30\% left polarised. 
The nominal beam parameter set was used to simulate the
beam energy spread and beamstrahlung.

The study has only considered $\stau$-pair production, other
open channels were considered as SUSY background.
This meant that the study was done without prior
knowledge of $\MXN{1}$, so that it has concentrated on
observables with low sensitivity to this parameter:
spectrum endpoints, cross sections and polarisation.
The expected effect of the uncertainty
on $\MXN{1}$ on the determination of $\mstau$ is
nevertheless quoted in a parametric form.

Throughout, it has been assumed that the knowledge of the
SM background will be good, so that any uncertainty on the
average SM background is small compared to $\sqrt{N_{SM}}$
at the final stage of event selection. The same was assumed
for the determination of the selection efficiency.
The SUSY background, on the other hand, has been assumed to be
poorly known. It has been assigned a relative error of 100~\%
in most of the cases studied, or at best to be equal to
the Poisson fluctuations in signal-free control
samples, typically of about the same size as the final 
signal sample under study.

The results on the study of the $\stone$ production for the
spectrum endpoint, cross section, and $\tau$ polarisation were:
\begin{align*}
E_{\tau,max} = &42.11^{+0.14}_{-0.12} ~\GeV \\
\frac{\delta\sigma}{\sigma} =& 3.1~\% \\
\mathcal{P}_{\tau} = &91 \pm 6 \pm 5 ~\mathrm{(bkg)} \pm 3 ~\mathrm{(SUSY~masses)}~\% ~(\pi ~\mathrm{channel}) \\
\mathcal{P}_{\tau} = &86 \pm 5~\%~(\rho ~\mathrm{channel})
\end{align*}
The endpoint could be used to determine $\mstone$, and assuming
$\MXN{1}$ has been measured to its nominal value ($97.7~\GeVcc$)
with an error of $\delta \MXN{1}$, it was found to be
$$
\mstone = 107.73^{+0.03}_{-0.05}\pm 1.1 \cdot \delta \MXN{1} ~\GeV ~(\mathrm{endpoint}).
$$
Also the cross section could be used to determine $\mstone$. However,
$E_{CMS}$= 500 $\GeV$ is much too far from the threshold for this to 
be competitive: In this case $\Delta(\mstone)/\mstone$ would be  2.1\%
assuming a known mixing angle. This should be compared to $\Delta(\mstone)/\mstone
\sim 1~\permil$ from the end-point, with no assumption on the mixing angle.

The $\tau$ polarisation had a lower statistical error in the $\rho$ channel.
However, this must be taken with caution, because there was 
a substantial amount
of remaining SM background from $\gamma\gamma$ processes. Due to lack of
simulation statistics, its contribution is poorly known.

The results on the study of the $\sttwo$ production for the
spectrum endpoint and cross section were:
\begin{align*}
E_{\tau,max} = &151.0^{+2.0}_{-1.7} ~\GeV \\
\frac{\delta\sigma}{\sigma} = &4.2~\%
\end{align*}
The endpoint value yielded
$$
\msttwo = 183 ^{+11}_{-5} \pm 18 \cdot \delta \MXN{1} ~\GeVcc  ~(\mathrm{endpoint}).
$$
For the $\sttwo$, $E_{CMS}$= 500 $\GeV$ is much more favourable for the determination of the
mass from the cross section: the 
expected uncertainty was $\Delta(\msttwo)/\msttwo$ = 2.4\%,
comparable to what could be obtained from the endpoint ($\sim$ 4\%).
Hence, the two could be combined to determine $\MXN{1}$, and
the error was found to be 1.7 $\GeVcc$, similar to what was found
in a separate analysis of $\smul$ using the same simulated sample.
However, this value assumes that the mixing angle is known.

The $\stau$ mixing angle has not been studied in this paper, because
the most sensitive process for its determination - $\stone\sttwo$ production -
should be studied below the $\sttwo\sttwo$ threshold to get a good
signal to background ratio.

In comparison with previous studies, e.g.~\cite{bib:diagonal}, 
\cite{bib:ulilowdm} and \cite{bib:stauNLC}, several new aspects have been
taken into account here. Most prominently, the smearing of four-vectors with
design goal resolutions has been replaced by a detailed simulation of the various
sub-detectors, including support structures, read-out, coolling etc. 
Further realism has been added by including not only background from Standard 
Model processes and from 
beamstrahlung pairs, but also from other (non-signal) SUSY processes, not
always taken into account in the previous studies. 
The consideration of these additional backgrounds required 
improvements of the $\tau$ reconstruction and of the signal selection cuts. 
After these efforts, the achieved precision on the $\mstone$ and on the $\tau$ 
polarisation is comparable to previous studies. Precise quantitative comparisons
would need to take into account the different SUSY scenarios and accelerator 
parameters which have been used. The expected precision for the $\sttwo$ mass and 
cross section have not been evaluated in either of the previous studies.

More specifically, a similar ``SPS1a inspired'' scenario 
with a slightly smaller mass difference between $\stone$ and $\XN{1}$ has been studied 
in~\cite{bib:ulilowdm} with a fast simulation of the TESLA detector, i.e. a 
predecessor to ILD. With an integrated luminosity of 200~fb$^{-1}$ and a beam 
polarisation of  $\mathcal{P}_{beam}(e^+,e^-) = (-0.6,+0.8)$ at a centre-of-mass energy 
of 400~\GeV, statistical precisions on $\mstone$ of 140~\MeVcc, 100~\MeVcc~and 100~\MeVcc ~have 
been achieved in the single $\pi$, $\rho$ and 3$\pi$ channels, respectively, excluding 
any contribution from the uncertainty of $\MXN{1}$. Combined, this corresponds to a 
precision of about 60~\MeVcc, quite similar to the $^{+30}_{-60}$~\MeVcc obtained here. 
The higher integrated luminosity assumed in the study presented here is compensated 
in~\cite{bib:ulilowdm} by a higher degree of positron polarisation and more optimal 
choice of the centre-of-mass energy, i.e. a higher cross section. Concerning the $\tau$ 
polarisation, a precision of 7~\% has been achieved in~\cite{bib:diagonal} from 
the $\rho$ channel. There, a 
scenario with a significantly larger mass difference has been studied using the JLC
fast detector simulation, assuming an integrated luminosity of 100~fb$^{-1}$, an electron 
beam polarisation of 0.95 and a centre-of-mass energy of 500~\GeV. In view of the differences
in beam parameters and in the SUSY scenario, this is in good agreement with the results
of this study.

Finally, it has to be pointed out that for many of the processes
studied in this work, running the accelerator at $E_{CMS}$= 500 $\GeV$
is not optimal. 
An upcoming study will treat the entire SPS1a' scenario as a whole,
including how to partition the luminosity in an optimal way,
and how to make use of non-$\stau$ channels to measure parameters -
notably $\MXN{1}$ - that were found to be hard to access
in the $\stau$ channels.

\section{Acknowledgements}
\label{sec:5}
We would like to thank the simulation production team,
in particular F. G\"ade, S. Aplin, J. Engels and I. Marchesini,
for their great effort to produce the large samples of events
used in this work. We would also like to thank T. Barklow for
producing the generated input files for the SM backgrounds.
The help from Z. W\c as in interfacing TAUOLA with Whizard
was much appreciated.

We acknowledge the support of the DFG through 
the SFB (grant SFB 676/1-2006) and 
the Emmy-Noether program (grant LI-1560/1-1). 
%

%
%
%
%
%
%
%
%
%
%

%
%

%
%

%
%
%
%
%
%
%
%
%
%
%
%
%
%
%
%
\newpage
\cleardoublepage
%
%
%
%

\end{document}

%% file: newcom.tex
\def\leqsim{\mathbin{\;\raise1pt\hbox{$<$}\kern-8pt\lower3pt\hbox{$\sim$}\;}}
\def\geqsim{\mathbin{\;\raise1pt\hbox{$>$}\kern-8pt\lower3pt\hbox{$\sim$}\;}}

\def\MXN#1{\mbox{$ M_{\tilde{\chi}^0_#1}                                $}}

\def\XP#1{\mbox{$ \tilde{\chi}^+_#1                                     $}}
\def\XM#1{\mbox{$ \tilde{\chi}^-_#1                                     $}}
\def\XPM#1{\mbox{$ \tilde{\chi}^{\pm}_#1                                $}}
\def\XMP#1{\mbox{$ \tilde{\chi}^{\mp}_#1                                $}}
\def\XN#1{\mbox{$ \tilde{\chi}^0_#1                                     $}}

\def\p#1{\mbox{$ \mbox{\bf p}_1                                         $}}

\newcommand{\smu}     {\mbox{$ \tilde{\mu}                                 $}}
\newcommand{\smur}    {\mbox{$ \tilde{\mu}_{\mathrm R}                     $}}
\newcommand{\smul}    {\mbox{$ \tilde{\mu}_{\mathrm L}                     $}}

\newcommand{\sel}     {\mbox{$ \tilde{e}                           $}}
\newcommand{\sell}    {\mbox{$ \tilde{e}_{\mathrm L}               $}}
\newcommand{\selr}    {\mbox{$ \tilde{e}_{\mathrm R}               $}}

\newcommand{\stau}    {\mbox{$ \tilde{\tau}                                $}}
\newcommand{\stone}   {\mbox{$ \tilde{\tau}_1                              $}}
\newcommand{\sttwo}   {\mbox{$ \tilde{\tau}_2                              $}}

\newcommand{\mstau}   {\mbox{$ M_{\tilde{\tau}}                            $}}
\newcommand{\mstone}  {\mbox{$ M_{\tilde{\tau}_1}                          $}}
\newcommand{\msttwo}  {\mbox{$ M_{\tilde{\tau}_2}                          $}}

\newcommand{\eeto}    {\mbox{$ {\, e}^+ {e}^- \to             $}}

\newcommand{\MeV}     {\mbox{$ {\mathrm{MeV}}                              $}}

\newcommand{\MeVcc}   {\mbox{$ {\mathrm{MeV}}                              $}}
\newcommand{\GeV}     {\mbox{$ {\mathrm{GeV}}                              $}}
\newcommand{\GeVc}    {\mbox{$ {\mathrm{GeV}}                              $}}
\newcommand{\GeVcc}   {\mbox{$ {\mathrm{GeV}}                              $}}

\newcommand{\ba}{\begin{array}}
\newcommand{\ea}{\end{array}}
\newcommand{\bc}{\begin{center}}
\newcommand{\ec}{\end{center}}
\newcommand{\be}{\begin{eqnarray}}
\newcommand{\eeq}{\end{eqnarray}}
\newcommand{\bes}{\begin{eqnarray*}}
\newcommand{\ees}{\end{eqnarray*}}
\newcommand{\Kz}{\ifmmode {\rm K^0_s} \else ${\rm K^0_s} $ \fi}
\newcommand{\Zz}{\ifmmode {\rm Z^0} \else ${\rm Z^0 } $ \fi}
\newcommand{\xxbar}{\ifmmode {\rm x\bar{x}} \else ${\rm x\bar{x}} $ \fi}
\newcommand{\rphi}{\ifmmode {\rm R\phi} \else ${\rm R\phi} $ \fi}

\def    \missEt      {\ifmmode{/\mkern-11mu E_t}\else{${/\mkern-11mu E_t}$}\fi}
\def    \missE       {\ifmmode{/\mkern-11mu E}\else{${/\mkern-11mu E}$}\fi}
\def    \missp       {\ifmmode{/\mkern-11mu p}\else{${/\mkern-11mu p}$}\fi}
\def    \misspt      {\ifmmode{/\mkern-11mu p_t}\else{${/\mkern-11mu p_t}$}\fi}